# Variability of $H_{\alpha}$ emission in young stellar objects in the cluster IC 348

E. H. Nikoghosyan and N. M. Azatyan [1]
Byurakan Astrophysical Observatory, 0213, Aragatsotn prov., Armenia

## Abstract

H$\alpha$ emission is one of the most prominent features of young stellar objects in the optical range, and importantly, the equivalent width of H$\alpha$ emission (EW(H$\alpha$)) is used to characterize an evolutionary stage of young stars. The aim of this work is to identify and study the stellar objects with variable EW(H$\alpha$) in the young stellar cluster IC 348. We performed photometric and slit-less observations at several epochs in order to reveal the variable objects. Significant variability of EW(H$\alpha$) was found in 90 out of 127 examined stars. From all epochs of observations, 32 objects were classified as CTT (classical T Tauri) and 69 as WTT (weak-line T Tauri) objects. The fraction of the variables in these samples is ~60%. We also identified 20 stellar objects, which showed not only a significant variability of the equivalent width, but which also change their apparent evolutionary stage (CTT ↔WTT). For 6 stars, H$\alpha$ line was observed both in emission and in absorption.

The analysis of data obtained over a wide wavelength range (from X-ray to mid-infrared) has shown that H$\alpha$ activity and the measure of its variability are in good agreement with the activity of stellar objects measured with its other parameters, such as X-ray radiation and the mass accretion rate. The EW(H$\alpha$) differs not only between objects at different evolutionary stages, but also between variable and non-variable objects. The variables in the CTT and WTT samples are more active than non- variables although they have almost the same evolutionary age. Another distinct difference between these variables and non-variables is their average masses. The variables from both CTT and WTT samples are noticeably more massive than non-variables. Our data confirm the assumption made for other star formation regions, that the decay of accretion activity occurs more slowly for more massive CTT objects. Apparently, a similar trend is also present in WTT objects, which are at a later stage of evolution. The variability of the stellar objects, which change their evolutionary classes (CTT ↔WTT), at least in a fraction of them, is due to the fact that they are close binaries, which affects and modulates their H$\alpha$ emission activity.

## 1 INTRODUCTION

The existence of a circumstellar disk and an envelope, a stellar wind and bipolar outflows determines the unique observational features of young stellar objects that distinguish them from other stellar population. These features occur over a wide wavelength range from X-ray to radio. During the evolution from protostars to disk-less pre-main stellar objects, the quantitative ratio of these features changes.

One of the most prominent properties of young stellar objects in the optical range is the presence of emission lines, in particular the tracer of ionized gas, $H_{\alpha}$ emission (6563 Å). Furthermore, the intensity of $H_{\alpha}$ emission, more precisely the equivalent width EW($H_{\alpha}$), can be used to characterize the evolutionary stage of young stars (Herbig & Bell, 1988). Low-mass young stellar objects are conditionally divided into two main groups based on their EW($H_{\alpha}$). The first group includes CTT objects, in which $H_{\alpha}$ emission is mainly produced as a result of accretion activity of the circumstellar disk. The second group includes WTT objects, in which $H_{\alpha}$ emission is mainly due to chromospheric solar like flares. Supposedly, these objects have a later evolutionary stage (Class III), while CTT objects belong to an earlier evolutionary stage (Class II). The widely used boundary value of EW($H_{\alpha}$) for these two groups is 10 Å. The classification proposed by White & Basri (2003) is also used to separate stellar objects according to their $H_\alpha$ activity. According to this classification, the boundary value of the EW($H_\alpha$) for WTT and CTT objects depends on their spectral classes (Sp). The boundary value changes from 3 Å for Sp earlier than K5 up to 40 Å for Sp later than M6.

[1] E. H. Nikoghosyan and N. M. Azatyan

Another distinguishing property of the young stellar objects is both photometric and spectroscopic variability. According to the current model a vital and central process in the formation of low mass stars is magnetospheric accretion where the stellar magnetic field threads the disk, and the material in the disk falls along the field lines onto the surface of the star (Koenigl, 1991). The accretion flows and shocks emit continuum emission from the ultraviolet (UV) to the infrared (IR), as well as a number of emission lines (e.g., $H_\alpha$, Ca II, He I), which, when observed are all found to be variable on time-scales from hours to years (e.g. Nguyen et al., 2009). Changes in the accretion rate lead to changes in the veiling in the stellar continuum. This can lead to measured changes in the EW of emission lines, including $H_\alpha$. Therefore, the variability in optical, ultraviolet, and infrared wavelengths, including the variability in emission lines has the potential to provide information about a region of the disk that is well below the resolution limit of current telescopes (e.g. Bouvier et al., 2013; Venuti et al., 2015; Flaherty et al., 2013; Costigan et al., 2014). Specifically, the $H_\alpha$ excess is often used as an indicator of mass accretion, and its intensity can be empirically related to $L_{acc}$ (e.g. Costigan et al., 2014).

The aim of this paper is to study the variability of EW($H_\alpha$). The target of the study is the relatively well investigated young (2 – 3 Myr) IC 348 stellar cluster (Stelzer et al., 2012). IC 348 is associated with the Perseus molecular cloud complex located at ∽300 pc from the Sun. The extinction ($A_v$) in the cluster ranges from 1 up to 10 mag, with a mean value of ∽3.5 mag (Luhman et al., 2003).

The pioneering study of Herbig (1954) discovered 16 stars with $H_\alpha$ emission in this cluster. Later, Herbig (1998) published a list of ∽100 emission stars and studied those in detail. Even then, the author, noted that many stellar objects in the cluster show a significant variation of the $H_\alpha$ emission. Later, the number of known $H_\alpha$ emitters was doubled (Luhman et al., 2003). In Nikoghosyan et al. (2015), it was shown that among the cluster members with $13.0 \leqslant$ R mag $\leqslant 19.0$ the percentage of emission stellar objects reaches 80%.

IC 348 was also extensively studied in near- and mid-infrared wavelengths (Lada et al., 2006; Muench et al., 2007; Currie & Kenyon, 2009). Using measurements of infrared excess between 3.6 and 8.0 $\mu$m, Lada et al. (2006) find that the total frequency of disk bearing stars in the cluster is ∽50% and only ∽30% of the cluster members are surrounded by optically thick, primordial disks. As concluded in Stelzer et al. (2012), the evolutionary stage of IC 348 cluster corresponds to the time when the structure of the disks of most young stellar objects changes from primordial, rather massive accretion disks to transitional and debris disks. Therefore, the cluster population represents the outcome of a recent star formation event.

IC 348 is also well studied in the X-ray regime. About 200 X-ray sources, which are associated with known cluster members (Preibisch & Zinnecker, 2002; Stelzer et al., 2012, and ref. herein), were detected. The detection rate of X-ray emitters is higher for diskless Class III stars.

This paper presents the results of optical observations and their analysis for more than 100 members of the IC 348 cluster. The paper is laid out as follows: Section 2 provides a description of the process of observations, in Section 3.1 we analyze the obtained observational data and the variability of the EW($H_\alpha$), in Section 3.2 we discuss the different parameters of $H_\alpha$ emitters, including optical, infrared, and X-ray data, as well as the masses, the evolutionary ages, and the mass accretion rates. The obtained results are discussed in Section 4.

**Table 1** Log of Observations

| Date | $H_\alpha$ slit-less (exp., sec) | $R_c$ (exp., sec) | $I_c$ (exp., sec) |
|---|---|---|---|
| 20.01.2009 | 2400 | 600 | 600 |
| 06.11.2009 | 2400 | 600 | 600 |
| 13.11.2010 | 1800 | – | – |
| 19.11 2016 | 1800 | 360 | 360 |

## 2 OBSERVATIONS AND DATA ANALYSIS

The optical observations were carried out at the prime focus of the 2.6 m telescope of the Byurakan Astrophysical observatory (BAO) using the multi-mode focal reducer SCORPIO and CCD-Detector EEV 42-40. The field and resolution of the obtained images are 14′ x 14′ and 0.38 ″/pix, respectively. During the observations, the seeing was ∽1.7 ″.

The main purpose of these observations was to identify the stellar objects with $H_\alpha$ emission and to determine its equivalent width. For this we used the slit-less method combination of the grism, working in the 5500-7500 Å spectral range, with the narrow-band $H_\alpha$ interference filter ($\lambda_c$ = 6560 Å, $\Delta\lambda$ = 85 Å). The spectral resolution was 0.86 Å/pix. At an exposure of 1800 sec, the photometric limit of the objects, for which $H_\alpha$ emission was detected, is R $\leqslant$ 19.0 mag. To study the variability of the $H_\alpha$ emission, the observations were carried out at several epochs. The epochs, as well as the exposure time of the observations are shown in Table 1.

The equivalent widths of $H_\alpha$ line were measured using the MIDAS software. The measurement errors were determined by the formula proposed in Vollmann & Eversberg (2006):

$$\sigma(W_\lambda) = \sqrt{1 + \overline{F_c}/\overline{F}}(\Delta\lambda - W_\lambda)/(S/N), \qquad (1)$$

where $\overline{F_c}$ is the average level of the continuum, $\overline{F}$ is the flux in the spectral line, and S/N is the signal-to-noise ratio. On the average, for objects with $R < 17.0$ the measurement errors are 30% of the EW($H_\alpha$), and for fainter objects the errors increase up to 40%.

In conjunction with spectral observations, the photometric observations in R and I bands were also carried out (see Table 1). For photometric observations, the Cousins $R_c$ and $I_c$ filters were used. For calibration of the photometric data, the magnitudes of the stars in the NGC 7790 and NGC 2264 R standard fields were used. The R and I stellar magnitudes were determined using the APPHOT and DAOPHOT packages of the IRAF software. The errors of photometric measurements do not exceed 0.03 mag.

In addition to the data of our observations, the values of EW($H_\alpha$), as well as $R_c$ and $I_c$ stellar magnitudes, taken from Herbig (1998), Luhman (1999), and Luhman et al. (2003), were used.

Besides optical data, we used 2MASS JHK near-infrared magnitudes, *Spitzer* IRAC mid-infrared magnitudes of [3.6], [4.5], [5.8], [8.0], [24] bands from Lada et al. (2006), as well as the X-ray luminosities from Preibisch & Zinnecker (2002) and Stelzer et al. (2012).

## 3 RESULTS

### 3.1 $H_\alpha$ emission

#### *3.1.1 Variability*

The area of our observations, which is shown on Fig. 1, covers the central region of IC 348 cluster. In this region, there are 127 stellar objects, for which we made two or more measurements of EW($H_\alpha$). The spectral (EW($H_\alpha$)) and photometric (R and I mag) data, obtained at different epochs are presented in Table 2, which includes the object numbers in Luhman et al. (2003) and Flaherty et al. (2013) (N1 and N2, respectively), coordinates (RA and Dec), as well as $R_c$, $I_c$ magnitudes and the values of EW($H_\alpha$) obtained on the 2.6 m telescope of BAO and taken from Herbig (1998), Luhman (1999), and Luhman et al. (2003). The negative values of EW($H_\alpha$) correspond to an absorption. One stellar object (M185) was identified as an IC 348 cluster member in Muench et al. (2007) and was not considered in Luhman et al. (2003) and Flaherty et al. (2013). The $R_c$ and $I_c$ magnitudes of two close binaries with numbers 39, 40 and 55, 56 were measured together.

For determining the variability of EW($H_\alpha$), we used the parameter $f_v$ (variability fraction), which, as a matter of fact, is the ratio of the standard deviation of the

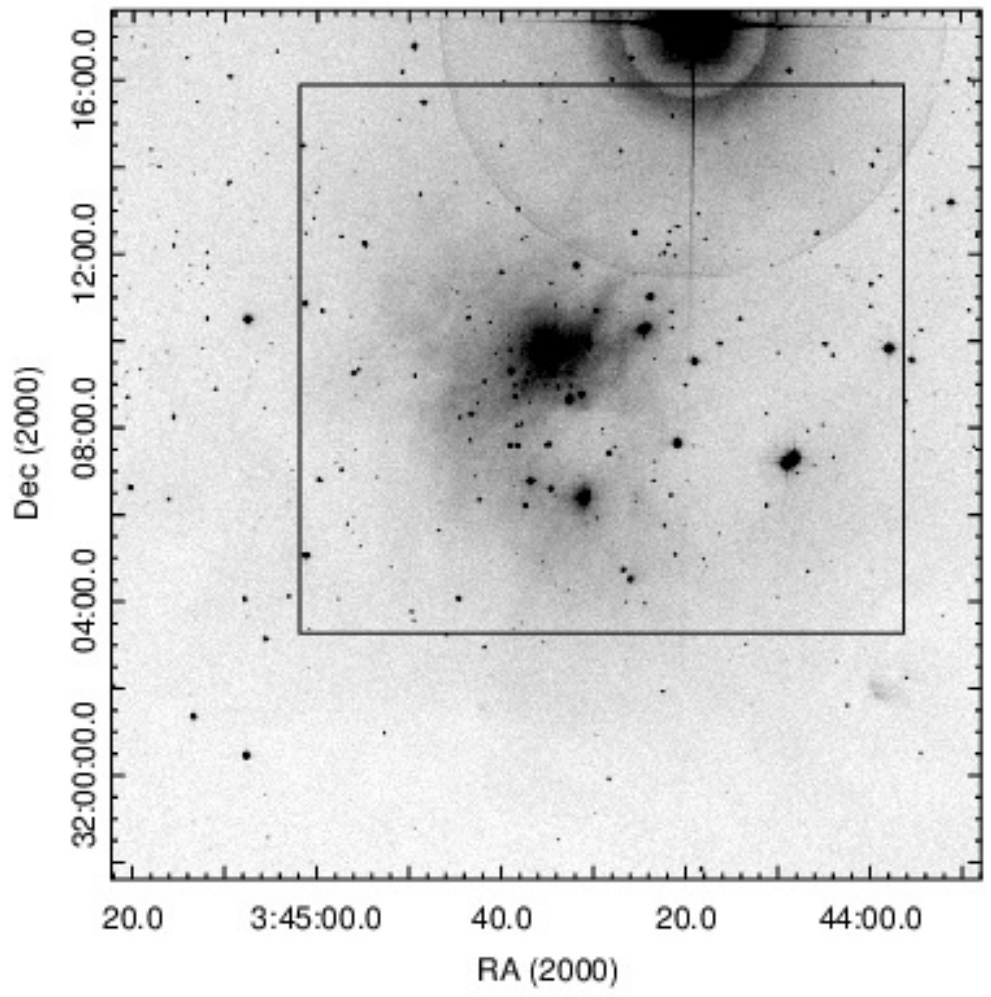

**Figure 1.** DSS2 R image of IC 348 cluster. The field of observation is shown by the box.

average value of the EW($H_\alpha$):

$$f_v = \sqrt{\frac{1}{n}\sum_{i=1}^{n}(x_i - \overline{x})}/\overline{x}, \qquad (2)$$

where n is the number of the measurements, $x_i$ is the EW($H_\alpha$) measured on different epochs of the observations, and $\overline{x}$ is the average value of EW($H_\alpha$). The values of both $f_v$ and average EW($H_\alpha$) for every stellar object are presented in Table 4. In this table, the averaged over all epochs $R_c$ mag (<R>) are also presented. Taking into account the errors of EW($H_\alpha$), we refer to as variables those objects with <R> brighter than 17.0 mag, for which $f_v > 0.3$, as well as the objects with <R> fainter than 17.0 mag, for which $f_v > 0.4$. In total, 90 out of 127 stars show a variability with respect to EW($H_\alpha$). We would like to note that the existence of a significant number of stars with variable $H_\alpha$ emission in this young stellar cluster was already indicated in the previous works (Herbig, 1998; Stelzer et al., 2012).

To check the selectivity of the variability fraction ($f_v$) with respect to the EW($H_\alpha$) and the brightness in the R band (R band, since $H_\alpha$ emission belongs to this spectral range), we used the linear regression. If the slope of the linear fitting is significantly different from zero, then the two variables are correlated. The distributions of the variability fraction ($f_v$) relative to the averaged over all epochs of the observations EW($H_\alpha$) and R band mag are presented in Fig. 2. The graphs in the Fig. 2 clearly show that there is practically no dependence between these parameters. The slopes and goodness of the linear fitting (R-square) for both distributions ($f_v$ vs. <EW($H_\alpha$)> and $f_v$ vs. <R>) are very small and equal to 0.0002 and 0.0011, as well as -0.0104 and 0.004, respectively. Therefore, the effect of selectivity with respect to these parameters on the result, is insignificant.

**Table 2** Data of observation

| N1 | N2 | RA (2000) | Dec (2000) | Herbig (1998), Oct. 1994 R$_c$ | I$_c$ | EW(H$_\alpha$) | Luhman et al. (2003), Dec. 1998[1] R$_c$ | I$_c$ | EW(H$_\alpha$) | 20.01.2009 R$_c$ | I$_c$ | EW(H$_\alpha$) | 06.11.2009 R$_c$ | I$_c$ | EW(H$_\alpha$) | 13.11.2010 EW(H$_\alpha$) | 18.11.2016 R$_c$ | I$_c$ | EW(H$_\alpha$) | 19.11.2016 R$_c$ | I$_c$ | EW(H$_\alpha$) | 21.11.2016 R$_c$ | I$_c$ | EW(H$_\alpha$) |
|---|---|---|---|---|---|---|---|---|---|---|---|---|---|---|---|---|---|---|---|---|---|---|---|---|---|
| 190 | 204 | 3 43 55.3 | 32 07 53 | 18.49 | 15.93 | 40 | | 16.08 | 34 | 18.21 | 16.23 | 72 | | 16.23 | | | | | | | | | | | 38 |
| 119 | 122 | 3 43 56.2 | 32 08 36 | 15.89 | 14.38 | | | 14.65 | 4 | 15.99 | 14.74 | 3 | | 14.74 | | | | | | 16.27 | 14.63 | 3 | | | 3 |
| 59 | 59 | 3 43 58.9 | 32 11 27 | 16.09 | 14.42 | 28 | | 14.42 | 27 | 16.06 | 14.62 | 21 | | 14.62 | | 65 | | | | 15.82 | 14.19 | 24 | 15.85 | 14.29 | |
| | 103 | 3 43 59.1 | 32 14 21 | | | 14 | | | | 17.40 | 16.19 | 12 | | 16.19 | | | | | | | | 36 | 17.09 | 15.70 | |
| 78 | 79 | 3 43 59.7 | 32 14 03 | | | | | 14.05 | 2 | 15.66 | 14.57 | 8 | | 14.57 | | 9 | | | | | | | 15.67 | 14.21 | |
| 143 | 147 | 3 44 04.1 | 32 07 17 | 16.49 | 14.85 | | | 15.01 | 3 | 16.38 | 14.92 | | | 14.92 | | 4 | 16.69 | 14.99 | | 16.58 | 14.90 | 7 | 16.51 | 14.69 | 2 |
| 110 | 112 | 3 44 04.2 | 32 13 50 | | | | | 14.87 | 5 | 17.20 | 15.27 | | | 15.27 | | 18 | 17.50 | 15.05 | | 17.36 | 15.03 | 14 | 17.09 | 14.94 | |
| 52 | 52 | 3 44 05.0 | 32 09 54 | 14.02 | 12.95 | | 14.09 | 13.02 | | 14.09 | 13.15 | -1 | 14.84 | 13.15 | 1 | 2 | 14.35 | 13.14 | | 14.21 | 13.06 | | 14.18 | 12.43 | |
| 152 | 158 | 3 44 06.1 | 32 07 07 | 18.6 | 16.1 | | | 16.36 | 6 | 17.51 | 15.70 | | | 15.70 | | | 17.70 | 15.88 | 4 | 17.90 | 15.75 | | 17.61 | 15.76 | |
| 129 | 132 | 3 44 06.8 | 32 07 54 | 17.37 | 15.13 | 41 | | 15.31 | 10 | 16.99 | 15.20 | 31 | 17.72 | 15.20 | 29 | 50 | 17.53 | 15.32 | 19 | 17.23 | 15.18 | 24 | 17.26 | 15.23 | 14 |
| 168 | 177 | 3 44 07.5 | 32 04 09 | 17.9 | | 14 | | 15.59 | 7 | 17.62 | 15.63 | 6 | | 15.63 | | | 17.83 | 15.61 | | 17.66 | 15.53 | 7 | 17.75 | 15.58 | 15 |
| 142 | 146 | 3 44 10.1 | 32 04 04 | 18.2 | | 85 | | 15.89 | 110 | 18.13 | 15.87 | 78 | | 15.87 | | | 18.21 | 15.89 | 43 | 18.28 | 15.77 | 91 | 18.27 | 15.87 | 82 |
| 134 | 138 | 3 44 11.2 | 32 08 16 | 17.37 | 14.75 | | | 15.12 | 11 | 17.05 | 14.91 | 5 | 17.59 | 14.91 | | 17 | 17.21 | 15.05 | 9 | 17.23 | 14.96 | 8 | 17.32 | 15.05 | 12 |
| 94 | 95 | 3 44 11.3 | 32 06 12 | 15.63 | 14.04 | 5 | | 14.32 | 2 | 15.64 | 14.15 | 3 | 16.26 | 14.15 | | 2 | 15.96 | 14.31 | 1 | 15.78 | 14.19 | 2 | 14.99 | 14.25 | 5 |
| 232 | 255 | 3 44 12.8 | 32 10 55 | 19.27 | 16.77 | | 18.86 | 16.79 | 6 | 18.81 | 16.73 | 9 | 19.08 | 16.73 | | | 19.44 | 16.92 | | 18.97 | 16.81 | | 19.00 | 16.75 | |
| 139 | 143 | 3 44 17.8 | 32 04 48 | 18.26 | 15.69 | 5 | | 15.78 | 7 | 17.78 | 15.50 | | 18.07 | 15.50 | | | 18.19 | 15.63 | | 17.94 | 15.56 | | 17.89 | 15.58 | 6 |
| 83 | 84 | 3 44 17.9 | 32 12 20 | | | | | 13.87 | 6 | 15.46 | 13.90 | 4 | 15.73 | 13.90 | 5 | 10 | 15.57 | 13.86 | 4 | 15.51 | 13.85 | 2 | 15.51 | 13.87 | 4 |
| 163 | 171 | 3 44 18.1 | 32 10 53 | 19.47 | 17.83 | 16 | 19.88 | 18.24 | 30 | 17.44 | 15.93 | 5 | | 15.93 | | | | | | 19.67 | 18.12 | 33 | 19.28 | 18.20 | |
| 29 | 29 | 3 44 18.2 | 32 04 57 | 17.34 | 15.39 | 11 | | 15.37 | | 17.41 | 15.30 | 4 | 17.90 | 15.30 | | | 17.40 | 15.26 | 10 | 17.40 | 15.18 | 10 | 17.36 | 15.23 | 17 |
| 149 | 155 | 3 44 18.2 | 32 09 59 | 18.5 | 15.96 | 8 | | 15.74 | 6 | 18.35 | 15.36 | -4 | 18.14 | 15.36 | 4 | 17 | 18.01 | 15.64 | | 17.76 | 15.49 | | 17.81 | 15.54 | 6 |
| 195 | 211 | 3 44 18.3 | 32 07 33 | 18.26 | 15.99 | 5 | | 16.04 | 5 | 17.92 | 15.76 | | 17.91 | 15.76 | 14 | | 18.35 | 15.95 | | 18.08 | 15.89 | | 17.96 | 15.84 | 9 |
| 130 | 133 | 3 44 18.6 | 32 12 53 | | | | 18.56 | 16.43 | 115 | 18.78 | | 41 | | | 193 | 108 | 18.51 | 17.16 | | 18.27 | 16.57 | 37 | 19.04 | 16.95 | 40 |
| 228 | 251 | 3 44 19.20 | 32 06 00 | 19.48 | 16.85 | 9 | 19.16 | 16.91 | 10 | 19.05 | 16.65 | | | 16.65 | | | 19.35 | 16.78 | | 19.30 | 16.84 | | 19.33 | 16.82 | |
| 89 | 90 | 3 44 19.2 | 32 07 35 | 16.72 | | 4 | | 14.78 | 10 | 17.03 | 14.54 | 8 | | 14.54 | 8 | | 16.70 | 14.59 | | 16.38 | 14.48 | 6 | 16.20 | 14.42 | 4 |
| 271 | 304 | 3 44 19.6 | 32 06 46 | | | | 17.77 | 16.04 | 28 | | 15.55 | 11 | | 15.55 | | 33 | 17.69 | 15.66 | | 17.82 | 15.77 | 10 | 17.70 | 15.72 | 7 |
| 166 | 174 | 3 44 20.0 | 32 06 46 | 17.72 | 15.72 | 5 | | 15.81 | 9 | 17.55 | 15.55 | 2 | | 15.55 | | 11 | 17.68 | 15.66 | | 17.82 | 15.77 | 9 | 17.70 | 15.72 | 11 |
| 108 | 110 | 3 44 20.2 | 32 08 56 | 16.69 | 14.77 | 47 | | 14.84 | | 16.45 | 14.58 | 40 | | 14.58 | 36 | 75 | 16.77 | 14.75 | 65 | 16.64 | 14.69 | 40 | 16.61 | 14.71 | 41 |
| 209 | 230 | 3 44 20.3 | 32 05 44 | 19.51 | 16.94 | | 19.31 | 17.04 | 4 | 19.20 | 16.69 | 1 | | 16.69 | | | 19.48 | 16.82 | | 19.41 | 16.72 | | 19.02 | 16.75 | 8 |
| 102 | 104 | 3 44 21.3 | 32 05 02 | 16.69 | 14.95 | 30 | | 15.19 | 8 | 16.67 | 14.81 | 8 | 17.03 | 14.81 | | | 16.89 | 14.91 | 3 | 16.83 | 14.92 | 19 | 16.77 | 14.91 | 6 |
| 109 | 111 | 3 44 21.3 | 32 11 56 | 16.4 | 14.53 | | | 14.54 | | 16.31 | 14.58 | | 16.46 | 14.58 | 1 | 3 | 16.48 | 14.63 | | 16.45 | 14.61 | 1 | 16.33 | 14.59 | 5 |
| 167 | 176 | 3 44 21.3 | 32 12 37 | | | | | | 6 | 20.56 | | 31 | 20.91 | | 25 | 9 | | 17.95 | | 17.47 | 15.50 | | | 15.45 | |
| M185 | | 3 44 21.57 | 32 15 10 | | | | | | | 17.29 | 15.35 | 39 | 17.26 | 15.35 | 17 | 61 | | | | | | | 16.98 | 15.12 | |
| 101 | 102 | 3 44 21.6 | 32 10 17 | 16.21 | 14.56 | 2 | | 14.57 | | 15.98 | 14.47 | 2 | 16.36 | 14.47 | 8 | 4 | 16.38 | 14.57 | | 16.24 | 14.51 | 3 | 16.13 | 14.48 | 5 |
| 38 | 38 | 3 44 21.6 | 32 10 38 | 16.57 | 14.94 | 48 | | 14.99 | | 16.70 | 14.95 | 12 | 17.31 | 14.95 | 65 | 47 | 16.54 | 14.68 | 12 | 16.77 | 14.51 | 13 | 16.62 | 14.78 | 22 |
| 107 | 109 | 3 44 21.7 | 32 06 25 | 16.23 | 14.59 | | | 14.56 | 4 | 16.00 | 14.27 | 2 | 16.33 | 14.27 | 2 | 2 | 16.41 | 14.47 | | 16.18 | 14.42 | 4 | 16.12 | 14.43 | 6 |
| 148 | 153 | 3 44 21.8 | 32 12 31 | | | | | 14.92 | 4 | 16.72 | 14.94 | 1 | 17.00 | 14.94 | 2 | 4 | 16.88 | 14.99 | | 16.88 | 14.99 | 2 | 16.81 | 14.97 | 2 |
| 85 | 86 | 3 44 21.9 | 32 12 12 | | | | | 14.27 | 6 | 16.15 | 14.22 | | 16.32 | 14.22 | 7 | 9 | 16.36 | | | 16.34 | 14.31 | 4 | 16.23 | 14.27 | 3 |
| 57 | 57 | 3 44 22.3 | 32 05 43 | 16.67 | 15.19 | 34 | 16.71 | 15.23 | 51 | 15.89 | 15.23 | 27 | 16.41 | 15.23 | | | 16.87 | 15.19 | 39 | 16.97 | 15.34 | 45 | 16.77 | 15.20 | 47 |
| 91 | 92 | 3 44 22.30 | 32 12 01 | | | | | 14.47 | 90 | 16.04 | 14.35 | 75 | 16.26 | 14.35 | 122 | 89 | 16.23 | 14.43 | 55 | 16.27 | 14.49 | 32 | 16.13 | 14.43 | 56 |
| 103 | 105 | 3 44 23.0 | 32 11 57 | 16.04 | 14.4 | | | 14.33 | 2 | 15.93 | 14.32 | 4 | 16.15 | 14.32 | 5 | 2 | 16.10 | 14.34 | | 16.14 | 14.37 | 5 | 16.11 | 14.34 | 1 |
| 179 | 192 | 3 44 23.6 | 32 09 34 | 17.96 | 15.66 | 28 | | 15.74 | 7 | 17.57 | 15.48 | 5 | 17.78 | 15.48 | | 7 | 17.81 | 15.56 | | 17.84 | 15.60 | 3 | 17.70 | 15.59 | 3 |
| 82 | 83 | 3 44 23.7 | 32 06 46 | 15.82 | 14.08 | 3 | | 14.2 | 4 | 15.76 | 13.94 | 3 | 15.93 | 13.94 | 8 | 2 | 16.00 | 14.13 | | 15.81 | 14.04 | 7 | 15.85 | 14.10 | 7 |
| 105 | | 3 44 24.6 | 32 03 57 | 16.9 | | 3 | | 15.36 | | 17.13 | 15.23 | | 17.29 | 15.23 | | | 17.36 | 15.26 | 8 | 17.02 | 15.07 | 5 | 17.04 | 15.13 | 4 |
| 116 | 119 | 3 44 25.3 | 32 10 13 | 17.76 | 15.39 | 51 | | 15.31 | 55 | | | | | | | | | | | | | | | | |
| 55 | 56 | 3 44 25.5 | 32 11 31 | 17.03 | 15.13 | | | 15.98 | | 16.42 | 14.59 | 25 | 16.75 | 14.59 | | | 16.61 | 14.69 | 30 | 16.68 | | 49 | 16.45 | | 59 |
| 56 | 55 | 3 44 25.5 | 32 11 31 | 18.31 | 16.07 | 51 | | 16.14 | | | | 26 | | | 54 | 74 | | | 58 | | | | | | |
| 87 | 88 | 3 44 25.6 | 32 06 17 | 18.31 | 16.07 | | | 15.98 | 3 | 17.65 | 15.47 | 1 | 17.84 | 15.47 | | | 18.08 | 15.65 | | 17.80 | 14.57 | 6 | 17.82 | 15.67 | 11 |
| 60 | 60 | 3 44 25.6 | 32 12 30 | | | | | 13.4 | 3 | 15.03 | 13.63 | | 15.08 | 13.63 | 3 | 2 | 15.05 | 13.62 | 2 | 15.21 | 13.76 | 2 | 14.97 | 13.65 | 3 |
| 5 | 5 | 3 44 26.0 | 32 04 30 | 13.72 | 12.37 | 23 | 13.79 | 12.44 | 5 | 13.91 | 12.45 | 32 | 14.13 | 12.45 | | | 14.05 | 12.45 | 7 | 13.94 | 12.38 | 18 | 13.89 | 12.41 | 16 |
| 58 | 58 | 3 44 26.6 | 32 03 58 | 16.2 | | 11 | | 14.03 | 10 | 16.22 | 14.08 | 12 | 16.36 | 14.08 | | | 16.13 | 13.94 | 10 | 16.06 | 15.06 | 11 | 16.05 | 14.76 | 11 |
| 71 | 72 | 3 44 26.7 | 32 08 20 | 18.01 | 16 | 42 | | 16.01 | 75 | 17.66 | 15.66 | 35 | 17.77 | 15.66 | 45 | 70 | 17.69 | 15.65 | | 17.76 | 15.68 | 52 | 17.66 | 15.68 | 59 |
| 65 | 65 | 3 44 27.0 | 32 04 44 | 14.75 | 13.48 | 3 | 14.96 | 13.69 | 4 | 15.19 | 13.72 | 2 | 15.24 | 13.72 | | | 15.02 | 13.55 | 3 | 14.91 | 13.49 | 5 | 14.86 | 13.51 | 3 |
| 159 | 165 | 3 44 27.2 | 32 10 38 | 18.61 | 16.23 | 26 | | 15.88 | 100 | 19.03 | 16.29 | 83 | 18.56 | 16.29 | 68 | 54 | 18.49 | 16.20 | 23 | 18.83 | 16.37 | 71 | 18.92 | 16.58 | 98 |
| | 113 | 3 44 27.2 | 32 14 21 | | | | | | | 16.97 | 15.39 | 35 | 17.00 | 15.39 | 27.1 | 38 | | 15.73 | | | | 27 | 17.56 | 15.69 | |
| 227 | 250 | 3 44 27.3 | 32 07 18 | 18.91 | 16.67 | 6 | 18.73 | 16.72 | 5 | 18.42 | 16.41 | 7 | 18.55 | 16.41 | | | 18.78 | 16.53 | | 18.62 | 16.48 | | 18.62 | 16.50 | 9 |
| 77 | 78 | 3 44 27.9 | 32 07 32 | 15.79 | 14.06 | 3 | | 14.13 | 4 | 15.60 | 13.87 | 5 | 15.72 | 13.87 | 2 | 2 | 15.79 | 13.94 | 3 | 15.76 | 13.93 | 6 | 15.68 | 13.94 | 2 |
| 238 | 264 | 3 44 28.0 | 32 05 20 | 19.53 | 17.11 | 30 | 19.28 | 17.18 | 14 | 19.33 | 16.97 | | 18.81 | 16.97 | | | 19.55 | 16.96 | | 19.03 | 16.89 | | 18.60 | 16.99 | 23 |
| 76 | 77 | 3 44 28.1 | 32 16 00 | | | | | 14.11 | 5 | 16.45 | 14.80 | 3 | 16.34 | 14.80 | | | | | | | | | | | |
| 62 | 62 | 3 44 28.5 | 32 07 22 | 14.54 | 13.4 | | 14.67 | 13.53 | 1 | 14.42 | 13.19 | 2 | 14.51 | 13.19 | -1 | 3 | 14.47 | 13.22 | -1 | 14.45 | 13.20 | | 14.31 | 13.15 | 2 |
| 186 | 200 | 3 44 29.1 | 32 07 57 | 17.82 | 15.71 | 4 | | 15.79 | | 17.48 | 15.43 | | 17.53 | 15.43 | 7 | 7 | 17.93 | 15.62 | | 17.64 | 15.54 | 16 | 17.50 | 15.58 | |
| 37 | 37 | 3 44 29.7 | 32 10 40 | 15.7 | 14.26 | 145 | 15.54 | 14.1 | 104 | 15.68 | 14.13 | 71 | 15.59 | 14.13 | 109 | 146 | 15.81 | 14.15 | | 15.66 | 14.05 | 54 | 15.62 | 14.10 | 89 |
| 247 | 276 | 3 44 30.0 | 32 09 40 | 20.8 | 18.23 | 50 | 21.15 | 18.43 | 160 | 19.11 | 17.61 | | | 17.61 | | | 19.96 | | | | 18.09 | | | 18.01 | |
| 118 | 121 | 3 44 30.6 | 32 06 30 | 16.75 | 15 | | 16.43 | 15.07 | | | | 2 | | | | 5 | | | | 15.89 | 14.61 | 2 | 15.93 | 14.45 | |
| 198 | 217 | 3 44 31.0 | 32 05 46 | 19.28 | 16.74 | 6 | | 16.75 | 7 | | 16.07 | | | 16.07 | | | 18.73 | 16.54 | | 18.85 | 16.52 | | 18.86 | 16.52 | |
| 176 | 186 | 3 44 31.2 | 32 05 59 | 19.29 | 16.74 | 100 | | 18.28 | 200 | | | | | | | | 18.46 | | | 19.19 | 17.67 | | 19.42 | 18.04 | |
| 212 | 233 | 3 44 31.3 | 32 09 29 | | | | 19.21 | 16.75 | 10 | | 16.29 | 10 | | 16.29 | | | | | | 19.46 | 16.91 | | 18.73 | 16.44 | |
| 138 | 142 | 3 44 31.3 | 32 10 47 | 18.08 | 15.91 | | | 15.84 | 39 | 17.91 | 15.80 | 52 | 17.81 | 15.80 | 96 | 20 | 17.95 | 15.88 | | 18.19 | 15.86 | 23 | 18.02 | 15.89 | |
| 175 | 185 | 3 44 31.4 | 32 11 29 | 18.37 | 15.98 | | | 15.74 | 3 | 18.02 | 15.70 | 3 | 17.98 | 15.70 | 2 | 4 | | 15.73 | | 18.12 | 15.73 | | 18.01 | 15.70 | 5 |
| 187 | 201 | 3 44 31.7 | 32 06 53 | 18.57 | 16.02 | 9 | | 16.1 | 11 | | | | | | | | 18.38 | 15.91 | | 18.38 | 15.94 | 9 | 18.20 | 15.86 | |
| 223 | 246 | 3 44 32.4 | 32 03 27 | 20 | | 140 | 19.91 | 17.63 | 200 | 19.16 | | | | | | | | 17.61 | | | 17.35 | | | | 24 |
| 31 | 31 | 3 44 32.6 | 32 08 42 | 14.87 | 13.25 | 3 | | 13.25 | 6 | | | 1 | | | 1 | | | | | 14.13 | 12.81 | | | | |

**Table 2 – continued from previous page**

| N1 | N2 | RA (2000) | Dec (2000) | Herbig (1998), Oct. 1994 | | | Luhman et al. (2003), Dec. 1998[1] | | | 20.01.2009 | | | 06.11.2009 | | | 13.11.2010 | 18.11.2016 | | | 19.11.2016 | | | 21.11.2016 | | |
|---|---|---|---|---|---|---|---|---|---|---|---|---|---|---|---|---|---|---|---|---|---|---|---|---|---|
| | | | | $R_c$ | $I_c$ | EW($H_\alpha$) | $R_c$ | $I_c$ | EW($H_\alpha$) | $R_c$ | $I_c$ | EW($H_\alpha$) | $R_c$ | $I_c$ | EW($H_\alpha$) | EW($H_\alpha$) | $R_c$ | $I_c$ | EW($H_\alpha$) | $R_c$ | $I_c$ | EW($H_\alpha$) | $R_c$ | $I_c$ | EW($H_\alpha$) |
| 66 | 67 | 3 44 32.6 | 32 08 56 | 16.23 | 13.25 | 5 | | 14.32 | 5 | 16.34 | 14.15 | | 16.05 | 14.15 | 1 | | 16.24 | 14.22 | 1 | 16.21 | 14.22 | 1 | 16.12 | 14.21 | |
| 79 | 80 | 3 44 32.8 | 32 09 16 | 16.63 | 14.63 | 4 | | 14.63 | 5 | 16.23 | 14.52 | 3 | 16.31 | 14.52 | 5 | 5 | 16.50 | 14.52 | | 16.47 | 14.53 | 1 | 16.43 | 14.54 | 1 |
| 80 | 81 | 3 44 33.3 | 32 09 40 | 16.38 | 14.55 | | 15.9 | 14.55 | 5 | | | | | | 1 | 1 | | | | | | | 16.14 | | 1 |
| 61 | 61 | 3 44 34.0 | 32 08 54 | 14.65 | 13.45 | 2 | 14.89 | 13.69 | 1 | | | | | | 3 | 1 | 14.92 | 13.49 | | 14.88 | 13.49 | 2 | 14.78 | 13.46 | 2 |
| 202 | 222 | 3 44 34.1 | 32 06 57 | 19.81 | 17.01 | 15 | | 17.09 | 45 | | | | 19.56 | | | | | 16.77 | | 19.22 | 16.74 | 17 | 18.96 | 16.72 | 15 |
| 68 | 69 | 3 44 34.3 | 32 10 50 | 16.34 | 14.36 | 3 | | 14.36 | 5 | | | 2 | | | 5 | 5 | 16.48 | 14.34 | 2 | 16.30 | 14.30 | 2 | 16.21 | 14.28 | 1 |
| 160 | 166 | 3 44 34.4 | 32 06 25 | 18.52 | 15.93 | 8 | | 16.07 | 4 | | 15.69 | 2 | | 15.69 | | | 18.15 | 15.75 | | 18.14 | 15.73 | | 17.80 | 15.70 | |
| 126 | 129 | 3 44 34.8 | 32 11 18 | 16.93 | 14.85 | | | 14.84 | | 16.80 | 14.84 | -2 | 16.52 | 14.84 | 9 | -2 | 16.85 | 14.78 | 3 | 16.85 | 14.80 | | 16.76 | 14.79 | 3 |
| 45 | 45 | 3 44 34.9 | 32 06 34 | 14.44 | 13.26 | | 14.63 | 13.45 | 2 | | 13.17 | 3 | | 13.17 | 2 | | 14.45 | 13.10 | | 14.53 | 13.18 | 2 | 14.43 | 13.16 | 3 |
| 23 | 23 | 3 44 35.0 | 32 07 37 | | | | | | 0 | | | 2 | | | 1 | -1 | 14.11 | 12.85 | -1 | 14.10 | 12.68 | | 13.93 | 12.59 | 1 |
| 24 | 24 | 3 44 35.4 | 32 07 36 | | | 32 | | | 5 | 18.11 | | 13 | 18.21 | | 40 | 24 | | 16.41 | 10 | | | 9 | | | 12 |
| 135 | 139 | 3 44 35.5 | 32 08 56 | 18.67 | 16.07 | 50 | | 16.15 | 74 | | 15.67 | 8 | | 15.67 | | 8 | 18.41 | 15.83 | | 18.53 | 15.85 | | 17.93 | 15.89 | |
| 189 | 203 | 3 44 35.7 | 32 04 53 | 18.45 | 16.03 | 11 | | 16.1 | 7 | 18.30 | 16.01 | 7 | 18.15 | 16.01 | | | 18.28 | 15.89 | | 18.29 | 15.95 | 9 | 17.78 | 15.52 | 12 |
| 124 | 127 | 3 44 37.0 | 32 08 34 | 18.02 | 15.61 | 8 | | 15.66 | 30 | 17.56 | 15.96 | 8 | 17.39 | 15.96 | 5 | 28 | 17.70 | 15.83 | 8 | 17.70 | 15.45 | 11 | 17.67 | 15.41 | |
| 99 | 100 | 3 44 37.2 | 32 09 16 | 15.45 | 14.23 | | 15.63 | 14.41 | | 15.65 | 14.28 | 3 | 15.47 | 14.28 | 1 | 1 | 15.73 | 14.21 | | 15.58 | 14.18 | 3 | 15.51 | 14.20 | 6 |
| 75 | 76 | 3 44 37.4 | 32 09 01 | 16.28 | 14.64 | 7 | | 14.93 | | 16.41 | 14.78 | 3 | 16.59 | 14.78 | 6 | 2 | 16.47 | 14.60 | | 16.48 | 14.70 | 2 | 16.15 | 14.52 | 6 |
| 96 | 97 | 3 44 37.4 | 32 12 24 | 18.1 | | 10 | | 15.78 | 22 | 17.77 | 15.68 | 22 | 17.89 | 15.68 | 44 | | 17.95 | 15.77 | | 18.06 | 15.86 | 49 | 17.86 | 15.73 | 38 |
| 128 | 131 | 3 44 37.8 | 32 12 18 | 18.1 | | 6 | | 15.38 | 7 | 17.43 | 15.37 | 2 | 17.56 | 15.37 | 9 | 10 | 17.43 | 15.18 | | 17.52 | 14.29 | 12 | 17.30 | 15.17 | 13 |
| 30 | 30 | 3 44 37.9 | 32 08 04 | 15.72 | 14.08 | 68 | | 14.18 | 57 | 15.60 | 13.95 | 45 | 15.46 | 13.95 | 13 | 177 | 15.86 | 14.11 | 106 | 15.92 | 14.16 | 71 | 15.93 | 14.22 | 63 |
| 34 | 34 | 3 44 38.0 | 32 03 30 | | | | | 13.18 | | 14.68 | 13.44 | 8 | 14.92 | 13.44 | | | 15.40 | 13.86 | 29 | 15.27 | 13.82 | 36 | 15.57 | 14.17 | 25 |
| 158 | 164 | 3 44 38.0 | 32 11 37 | 17.77 | 15.57 | 10 | | 15.52 | 50 | 17.68 | 15.54 | 4 | 17.41 | 15.54 | | 56 | 17.63 | 15.46 | 11 | 17.62 | 15.49 | 25 | 17.48 | 15.41 | 5 |
| 33 | 33 | 3 44 38.4 | 32 07 36 | 14.37 | 13.07 | 2 | 14.56 | 13.26 | 4 | 14.56 | 12.99 | 3 | 14.46 | 12.99 | 5 | 1 | 14.59 | 13.08 | 3 | 14.47 | 13.01 | 3 | 14.63 | 13.17 | 2 |
| 120 | 123 | 3 44 38.4 | 32 13 00 | | | | | 14.57 | 2 | 16.19 | 14.83 | 2 | 15.89 | 14.83 | 2 | 2 | 16.11 | 14.60 | 5 | 16.17 | 14.69 | 3 | 16.04 | 14.60 | |
| 53 | 53 | 3 44 38.5 | 32 08 01 | 15.73 | 14.1 | 9 | | 14.24 | | 16.01 | 14.02 | 3 | 15.91 | 14.02 | 38 | 29 | 15.69 | 13.94 | 10 | 16.11 | 14.34 | 9 | 15.60 | 13.97 | 12 |
| 88 | 89 | 3 44 38.6 | 32 05 06 | 16.95 | 14.8 | 4 | | 14.88 | 4 | 16.96 | 14.84 | 9 | 16.91 | 14.84 | | | 16.86 | 14.65 | 7 | 16.84 | 14.69 | 4 | 16.85 | 14.73 | 5 |
| 95 | 96 | 3 44 38.7 | 32 08 57 | 16.18 | 14.37 | 3 | | 14.46 | 5 | 16.10 | 14.31 | 2 | 16.01 | 14.31 | 1 | | 16.22 | 14.30 | | 16.13 | 14.28 | 2 | 16.07 | 14.28 | 1 |
| 206 | 227 | 3 44 38.9 | 32 06 36 | | | | 18.87 | 16.6 | 10 | 19.03 | 16.40 | 12 | 18.52 | 16.40 | | | 18.99 | 16.37 | | 18.82 | 16.37 | | 18.63 | 16.33 | |
| 207 | 228 | 3 44 39.0 | 32 03 20 | 18.3 | | 98 | | 16.4 | 110 | 18.24 | 16.42 | 38 | 18.51 | 16.42 | | | 18.29 | 16.23 | | 18.35 | 16.38 | 53 | 18.77 | 16.52 | 49 |
| 32 | 32 | 3 44 39.2 | 32 07 35 | 14.43 | 13.03 | | 14.61 | 13.21 | | 14.56 | 12.99 | -1 | 14.83 | 12.99 | -1 | 1 | 14.59 | 12.97 | | 14.59 | 13.00 | | 14.50 | 12.98 | |
| 81 | 82 | 3 44 39.2 | 32 09 45 | 16.62 | 14.76 | 8 | | 14.76 | | 16.43 | 14.84 | 3 | 16.30 | 14.84 | 24 | 18 | 16.78 | 14.79 | 3 | 16.61 | 14.66 | 2 | 16.46 | 14.66 | |
| 197 | 216 | 3 44 39.4 | 32 10 08 | 18.4 | 16.18 | | | 16.06 | 5 | 17.99 | 16.03 | | 17.99 | 16.03 | | 3 | 18.08 | 15.94 | | 18.05 | 16.03 | | 18.21 | 16.01 | |
| 172 | 182 | 3 44 40.2 | 32 09 33 | 19 | 16.59 | 26 | 18.65 | 16.57 | 40 | 18.28 | 16.53 | 47 | 18.63 | 16.53 | | 51 | 19.21 | 16.55 | | 18.71 | 16.39 | | 19.23 | 16.46 | |
| 169 | 179 | 3 44 40.8 | 32 13 07 | 19 | | 9 | | 15.94 | 6 | 18.23 | 16.22 | 3 | 18.09 | 16.22 | | | 18.32 | 15.96 | | 18.38 | 16.16 | 6 | 18.05 | 15.96 | |
| 121 | 124 | 3 44 41.3 | 32 10 25 | 16.9 | 14.68 | 3 | | 14.69 | 4 | 16.77 | 14.75 | 5 | 16.65 | 14.75 | 7 | 7 | 16.80 | 14.60 | 7 | 16.79 | 14.63 | 8 | 16.84 | 14.60 | 5 |
| 173 | 183 | 3 44 41.4 | 32 13 09 | 18.3 | | 3 | | 15.68 | 5 | 17.89 | 15.84 | | 17.78 | 15.84 | 17 | | 17.83 | 15.66 | | 18.15 | 15.86 | | 17.97 | 15.69 | |
| 251 | 280 | 3 44 41.6 | 32 10 39 | 20.13 | 17.82 | 65 | 19.95 | 17.81 | 140 | 19.59 | 17.65 | | 19.14 | 17.65 | | | 20.52 | 17.66 | | 20.39 | 17.73 | | 19.77 | 17.61 | |
| 111 | 114 | 3 44 41.7 | 32 12 02 | | | | | 16.23 | | 18.50 | 16.20 | 20 | 18.28 | 16.20 | | 8 | 18.69 | 16.03 | | 18.65 | 16.08 | | 18.11 | 15.93 | |
| 39 | 39 | 3 44 42.0 | 32 09 00 | 16.83 | 14.85 | 32 | | 14.88 | 22 | 16.21 | 14.33 | 82 | 16.37 | 14.33 | 80 | 112 | 16.41 | | 17 | 16.54 | | 57 | 16.42 | | 63 |
| 40 | 40 | 3 44 42.1 | 32 09 02 | 17.03 | 15.16 | 74 | | 15.83 | 55 | | | 218 | | | 113 | 214 | | 14.24 | 36 | | 14.78 | 124 | | 14.33 | 96 |
| 122 | 125 | 3 44 42.6 | 32 06 19 | 15.02 | 13.89 | | 15.12 | 13.99 | 2 | 15.25 | 14.02 | 4 | 15.11 | 14.02 | | 1 | 15.12 | 13.83 | | 15.18 | 13.88 | 3 | 15.07 | 13.87 | 3 |
| 127 | 130 | 3 44 42.8 | 32 08 34 | 18.03 | 15.78 | 91 | | 15.95 | 40 | 18.20 | 15.81 | 79 | 17.79 | 15.81 | 54 | 29 | 17.81 | 15.53 | 38 | 17.78 | 15.57 | 91 | 17.71 | 15.22 | |
| 170 | 180 | 3 44 43.0 | 32 10 15 | | | | | 16.07 | 10 | 18.31 | 16.13 | | 18.30 | 16.13 | 6 | | 18.48 | 15.99 | | 18.48 | 16.04 | | 18.42 | 15.95 | |
| 49 | 49 | 3 44 43.5 | 32 07 43 | 16.71 | 14.86 | | 16.67 | 14.98 | | 16.82 | | 2 | | | 1 | 3 | 16.79 | 14.74 | | 16.83 | 14.71 | 3 | 16.69 | 14.75 | |
| 233 | 256 | 3 44 43.7 | 32 10 48 | 18.5 | 16.41 | 8 | | 16.4 | 9 | 18.64 | 16.57 | | 18.53 | 16.57 | | 1 | 18.79 | 16.43 | | 18.60 | 16.47 | 6 | 18.24 | 16.32 | 5 |
| 171 | 181 | 3 44 44.6 | 32 07 30 | 19.57 | 16.72 | 8 | 19.3 | 16.87 | 11 | 18.65 | 16.63 | | 18.91 | 16.63 | | | 18.84 | 16.49 | | 19.38 | 16.67 | | 19.14 | 16.53 | |
| 93 | 94 | 3 44 44.6 | 32 08 12 | 17.2 | 15.38 | 44 | | 15.73 | 26 | 16.88 | 15.29 | 31 | 16.98 | 15.29 | | 80 | 16.87 | 15.14 | 90 | 16.92 | 15.13 | 34 | 16.79 | 15.09 | 60 |
| 15 | 15 | 3 44 44.7 | 32 04 03 | | | | | 13.34 | 36 | 15.42 | 13.62 | 46 | 15.11 | 13.62 | | | 15.05 | 13.24 | | 15.26 | 13.49 | 49 | 15.18 | 13.46 | 39 |
| 141 | 145 | 3 44 44.8 | 32 11 06 | 17.22 | 15.28 | | | 15.28 | 3 | 17.42 | 15.50 | 8 | 17.13 | 15.50 | | | 17.44 | 15.23 | | 17.32 | 15.29 | 12 | 17.15 | 15.21 | 7 |
| 98 | 99 | 3 44 45.0 | 32 13 36 | 17.4 | | 7 | | 15 | 11 | 17.57 | 14.94 | 10 | 17.25 | 14.94 | | | 17.34 | 15.04 | | 17.73 | 15.35 | | 17.32 | 15.08 | |
| 147 | 151 | 3 44 48.8 | 32 13 22 | 16.5 | | 3 | | 14.94 | 14 | 17.29 | 15.72 | 6 | 16.75 | 15.72 | | | 16.97 | 15.10 | 8 | 17.21 | 15.46 | 8 | 16.95 | 15.20 | |
| 150 | 156 | 3 44 53.7 | 32 06 52 | 18.01 | 15.74 | | | 15.84 | 5 | 18.42 | 16.19 | 8 | 17.96 | 16.19 | | | 18.02 | 15.65 | | | | | 18.33 | 15.91 | |
| 47 | 47 | 3 44 55.6 | 32 09 20 | | | | | 14.28 | 2 | 16.33 | 14.67 | | 16.07 | 14.67 | | 1 | 16.03 | 14.08 | 1 | | | | | | |
| 154 | 160 | 3 44 56.1 | 32 05 56 | | | | | 14.91 | 5 | 17.53 | | 4 | 16.65 | | | | 16.60 | 14.84 | | | | | | | |
| 20 | 20 | 3 44 56.1 | 32 09 15 | | | | | 13.21 | 5 | 15.08 | 13.79 | 3 | 14.48 | 13.79 | | 1 | 14.60 | 13.08 | 5 | | | | | | |
| 181 | 194 | 3 44 59.8 | 32 13 32 | | | | | 16.21 | 80 | | | | | | 53 | | | | | | | | | | |
| 162 | 169 | 3 45 01.5 | 32 12 29 | | | | | 15.25 | 5 | | | 3 | 17.47 | | | | | | | | | | | | |

[1]The date refers only to EW($H_\alpha$). Photometric data were obtained at different epochs: $R_c$ mag in Sep. 1998 and $I_c$ mag in Sep. 1999.

*3.1.2 Evolutionary stage*

For the classification of the young stellar objects from their $H_\alpha$ activity (CTT with optically thick circumstellar disk and disk-less WTT), the criterion proposed by White & Basri (2003) has been used. According to that criterion, the boundary value of EW($H_\alpha$) for WTT and CTT objects depends on their spectral classes (Sp). The spectral types of the objects were taken from Luhman et al. (2003) and presented in Table 4.

We have defined the evolutionary classes for all objects from our sample for all epochs of the observations. The data of epochal observations show that 101 $H_\alpha$ emitters do not change their evolutionary classes. Of these, 32 were classified as CTT and 69 as WTT objects. We denoted these samples as CTT⇆CTT and WTT⇆WTT (hereafter CC and WW), respectively. The fraction of the objects with variable EW($H_\alpha$) among these two samples is almost the same (63% - 64%) as shown in Table 3, where not only quantitative relations of objects of different samples are given, but the average values of other parameters, which will be discussed below. Twenty stellar objects show not only significant variability of the EW($H_\alpha$), but also change their apparent evolutionary classes. These comprise the sample CTT⇆WTT (hereafter CW). It should be noted that their average variability fraction is much greater than that of other variables (CC and WW). Moreover, the values of $f_v$ of all CW objects are greater than the boundary value (0.3 or 0.4). For 6 stars, the $H_\alpha$ line was observed both in emission and in absorption.

## 3.2 Comparison with other parameters

One distinct advantage of studying the IC 348 cluster is the wealth of multi-wavelengths data that are available in the literature. It gives us a chance to carry out the comparative analysis of variability of $H_\alpha$ emission with other parameters, including optical and infrared photometric data, as well as of X-ray data. For this purpose, we used data from other works: [3.6], [4.5], [5.8], [8.0], [24] bands magnitudes and the $\alpha$ slopes of the infrared spectral energy distribution (SED) from Lada et al. (2006); information about the variability in the mid-infrared (MIR) range from Flaherty et al. (2013); $L_x$ luminosities from Preibisch & Zinnecker (2002) and Stelzer et al. (2012); the extinctions ($A_v$), the total luminosities (Lum), the effective temperature ($T_{ef}$), and spectral types (Sp) from Luhman et al. (2003), as well as the optical periods from Cohen et al. (2004) and Cieza & Baliber (2006). The following data are given in the Table 4: (1) and (2) - numeric identification of objects as in Luhman et al. (2003) and Flaherty et al. (2013), (3) - average values of EW($H_\alpha$), (4) - the variability fractions of EW($H_\alpha$), (5) - classifications of stellar objects from EW($H_\alpha$) in all epochs of observations, (6) - the variability of EW($H_\alpha$), (7) - (10) - average values of R and I magnitudes and their standard deviations, (11) - (13) - 2MASS JHK magnitudes, (14) - (18) - *Spitzer* IRAC magnitudes, (19) - $\alpha^{3-8\,\mu m}$ slopes of SEDs, (20) - variability of objects in MIR, (21) - $L_x$ luminosity, (22) - (25) - extinction, total luminosity, effective temperature, and spectral types, (26) - optical period (value of 0 corresponds to stars for which no periodical variability of brightness was detected).

*3.2.1 Optical range*

Using the average values of the R and I magnitudes and their standard deviations, we considered the relationship between the variability of the $H_\alpha$ emission and brightness. The results are presented in Fig. 3. The left panels show the distribution of the EW($H_\alpha$) variability fraction ($f_v$) versus the standard deviation (sd) of the R and I magnitudes. For better visibility, we have split the stellar objects from different samples into several bins with a nearly equal numbers (right panels).

According to the graphs in Fig. 3, we can see that the variable objects from the CC sample show correlation between the fraction of EW($H_\alpha$) variability and the standard deviation of the R mag. The same dependence, but not so well expressed, can be seen also for variable objects from the WW sample. There is no such correlation for the I band. For CW and non-variable stellar objects, the dependence between the variability of EW($H_\alpha$) and brightness is not observed in either photometric bands.

Such correlation for the objects from both CC and WW samples could be expected, since the $H_\alpha$ wavelength is within the R band. It should be noted, however, that since the width of R band is 1500 Å, and the equivalent width of the $H_\alpha$ line, even for CTT objects, does not exceed 150 Å, the effect of the variability of the $H_\alpha$ emission on the entire R band cannot be significant. It can be assumed that a certain relationship between the variability of the continual (R band) and emission ($H_\alpha$) radiation is due to the fact that they are caused by the same processes: accretion for CTT objects and chromospheric flare activity for WTT objects. Nevertheless, the analysis of spectral and photometric data obtained at different observation epochs showed that an increase in both R band brightness and EW($H_\alpha$) is in-phase only for 40% of CC, 60% of WW and 35% of CW variables. It should be noted that in the analysis of data, we used brightness fluctuations only if they exceeded the error of photometric measurements (0.03 mag).

We compared the objects from different samples relative to the results of optical photometric monitoring taken from Cohen et al. (2004) and Cieza & Baliber (2006). It should be noted that the data are given for ∽80% of the objects from our sample (see Table 4). Information on how many of the total number of objects studied in the sample show periodic variability is presented in Table 3.

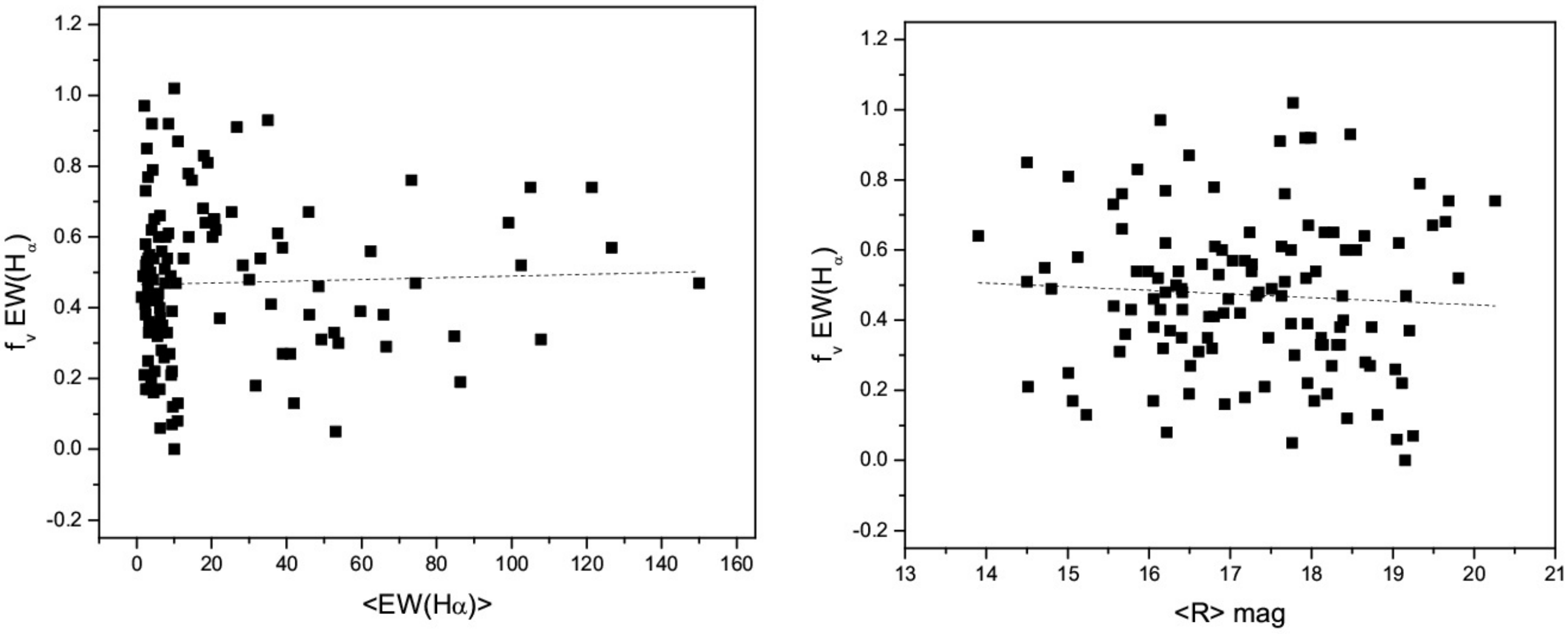

**Figure 2.** Distributions of the variability fraction ($f_v$) of EW($H_\alpha$) relative to the averaged over all epochs of the observations $<EW(H_\alpha)>$ (left panel) and $<R>$ band mag (right panel). The dashed lines are the result of the linear fitting. The selectivity of the variability fraction ($f_v$) with respect to the EW($H_\alpha$) and the brightness in the R band is not observed.

**Table 3** Averaged parameters of variable and non-variable stellar objects

| | CTTau⇆CTTau | | CTTau⇆WTTau | WTTau⇆WTTau | | WTTau⇆Abs |
|---|---|---|---|---|---|---|
| Data | Var. | Non-var. | Var. | Var. | Non-var. | |
| Number | 20 (63%) | 12 (37%) | 20 (100%) | 44 (64%) | 25 (36%) | 6 |
| $f_v$ of EW($H_\alpha$) | 0.43 ± 0.18 | | 0.62 ± 0.16 | 0.42 ± 0.21 | | |
| Periodic variability in optics | 8 of 15 | of 9 | 13 of 16 | 29 of 40 | 9 of 18 | 4 of 5 |
| J-K | 1.74 ± 0.34 | 1.59 ± 0.30 | 1.53 ± 0.59 | 1.26 ± 0.15 | 1.20 ± 0.11 | 1.21 ± 0.11 |
| $(J-K)_0$ | 1.04 ± 0.33 | 1.04 ± 0.21 | 1.06 ± 0.55 | 0.79 ± 0.16 | 0.78 ± 0.17 | 0.66 ± 0.14 |
| [3.6]-[8.0] | 1.44 ± 0.20 | 1.43 ± 0.29 | 1.08 ± 0.43 | 0.27 ± 0.55 | 0.19 ± 0.38 | 0.32 ± 0.35 |
| $([3.6]-[8.0])_0$ | 1.38 ± 0.23 | 1.38 ± 0.29 | 0.98 ± 0.52 | 0.23 ± 0.54 | 0.15 ± 0.39 | 0.28 ± 0.35 |
| [8.0]-[24] | 3.47 ± 1.06 | 3.41 ± 1.39 | 3.78 ± 1.58 | 5.36 ± 1.71 | 5.36 ± 1.65 | 3.70 ± 0.59 |
| $([8.0] - [24])_0$ | 3.28 ± 1.30 | 3.41 ± 1.39 | 3.78 ± 1.58 | 5.36 ± 1.71 | 5.36 ± 1.65 | 3.70 ± 0.59 |
| $-1.8 < \alpha < 0$ | 18 (95%) | 10 (91%) | 13 (%) | 6 (14%) | 1 (4%) | - |
| $-2.56 < \alpha < -1.8$ | 1 (10%) | 1 (8%) | 4 (22%) | 11 (25%) | 6 (24%) | 2 (33%) |
| $\alpha < -2.56$ | - | - | 1(6%) | 27 (61%) | 18 (72%) | 4 (67%) |
| IRAC var. | 6 of 6 | 2 of 3 | 6 of 8 | 11 of 22 | 3 of 8 | 1 of 4 |
| $A_v$ | 3.98 ± 1.44 | 3.50 ± 1.33 | 2.94 ± 1.22 | 2.84 ± 1.12 | 2.47 ± 1.26 | 3.20 ± 0.98 |
| Log ($L_x/L_{bol}$) | -4.16 ± 0.95 | -3.99 ± 0.62 | -3.75 ± 0.93 | -3.61 ± 0.50 | -3.70 ± 0.50 | -4.33 ± 0.69 |
| $M_*/M_{sun}$ (Siess et al., 2000) | 0.59 ± 0.57 | 0.29 ± 0.19 | 0.39 ± 0.41 | 0.36 ± 0.24 | 0.28 ± 0.21 | 0.74 ± 0.59 |
| Log (Age) (Siess et al., 2000) | 6.38 ± 0.30 | 6.62 ± 0.19 | 6.41 ± 0.18 | 6.47 ± 0.22 | 6.57 ± 0.14 | 6.46 ± 0.17 |
| $M_*/M_{sun}$ (Robitaille et al., 2007) | 0.81 ± 0.54 | 0.42 ± 0.33 | 0.43 ± 0.36 | 0.44 ± 0.28 | 0.30 ± 0.13 | 0.82 ± 0.36 |
| Log (Age) (Robitaille et al., 2007) | 6.39 ± 0.34 | 6.40 ± 0.28 | 6.34 ± 0.47 | 6.45 ± 0.34 | 6.48 ± 0.48 | 6.60 ± 0.29 |
| Log ($\dot{M}$) (Robitaille et al., 2007) | −8.17 ± 0.68 | −8.79 ± 0.88 | −8.76 ± 0.89 | −8.80 ± 0.93 | −8.95 ± 0.79 | −10.6 ± 1.17 |

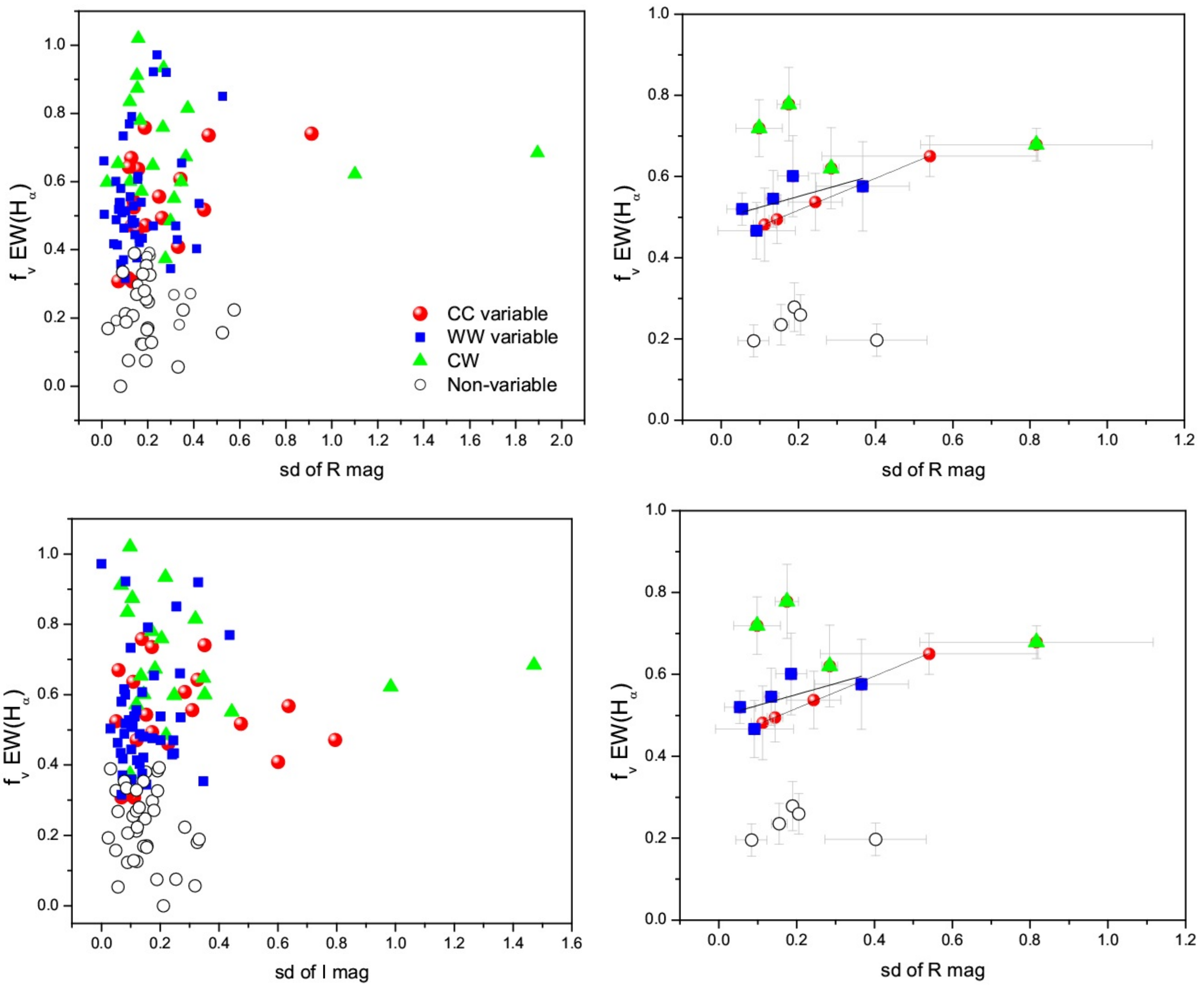

**Figure 3.** Variability fraction ($f_{\rm v}$) of EW(H$_\alpha$) vs. standard deviation (sd) of R (top panels) and I (bottom panels) magnitudes. Each bin on the right panels contains a nearly equal number of objects and the horizontal error bars represent the size of the bin. The vertical error bars represent the standard deviation of the $<f_{\rm v}>$. The "Non variable" sample includes objects from both CC and WW samples. On the top right panel the straight lines are the result of the linear fitting of objects from the CC and WW samples. The variable objects from CC and WW samples show certain correlation between the fraction of EW(H$_\alpha$) variability and the standard deviation of the R mag.

**Table 4** Main parameters

| N1 (1) | N2 (2) | EW (3) | $f_v$ (4) | Class. (5) | $V_{EW}$ (6) | <$R_c$> (7) | sd $R_c$ (8) | <$I_c$> (9) | sd $I_c$ (10) | J (11) | H (12) | K (13) | [3.6] (14) | [4.5] (15) | [5.8] (16) | [8.0] (17) | [24] (18) | $\alpha$ (19) | $V_{MIR}$ (20) | $\mathrm{Log}(L_x/L_{bol})$ (21) | $A_v$ (22) | Lum (23) | $T_{ef}$ (24) | Sp (25) | P(d) (26) |
|---|---|---|---|---|---|---|---|---|---|---|---|---|---|---|---|---|---|---|---|---|---|---|---|---|---|
| 5 | 5 | 18.4 | 0.64 | CC | y | 13.90 | 0.16 | 12.47 | 0.11 | 10.08 | 8.87 | 8.14 | 6.97 | 6.52 | 6.32 | 5.63 | 2.76 | -1.39 | y | -5.99 | 5.32 | 9.9 | 5520 | G8 | 6.5 |
| 29 | 29 | 9.1 | 0.49 | CC | y | 17.51 | 0.26 | 15.39 | 0.17 | 12.09 | 10.54 | 9.69 | 8.82 | 8.55 | 8.42 | 7.79 | 3.07 | -1.72 | y | -6.59 | 11.84 | 9.6 | 5945 | G1 | 0 |
| 30 | 30 | 73.3 | 0.76 | CC | y | 15.67 | 0.19 | 14.05 | 0.14 | 11.69 | 10.48 | 9.83 | 8.84 | 8.53 | 8.2 | 7.63 | 4.92 | -1.48 | y | -4.58 | 5 | 1.4 | 4060 | K7 | 8.2 |
| 37 | 37 | 107.8 | 0.31 | CC | y | 15.64 | 0.07 | 14.14 | 0.07 | 11.93 | 10.76 | 10.14 | 9.63 | 9.25 | 9.02 | 8.37 | 4.25 | -1.45 | | -4.64 | 3.58 | 0.76 | 3955 | K8 | 0 |
| 38 | 38 | 37.6 | 0.61 | CC | y | 16.81 | 0.34 | 15.00 | 0.28 | 12.49 | 11.28 | 10.62 | 9.44 | 8.79 | 8.63 | 8.18 | 5.35 | -1.50 | y | -3.97 | 5.64 | 0.79 | 4060 | K7 | 2.8 |
| 39 | 39 | 62.4 | 0.56 | CC | y | 16.65 | 0.25 | 14.69 | 0.31 | 11.78 | 10.66 | 10.13 | 9.38 | 8.95 | | 7.82 | 5.02 | -1.06 | | -3.69 | 3.3 | 0.37 | 3234 | M4,25 | 0 |
| 40 | 40 | 126.6 | 0.57 | CC | y | 17.03 | | 15.18 | 0.64 | | | | | | | | | | | -3.61 | 4.82 | 0.28 | 3488 | M2,5 | |
| 55 | 56 | 35.8 | 0.41 | CC | y | 16.80 | 0.33 | 15.56 | 0.60 | 12.82 | 11.59 | 11.07 | 9.54 | 9.12 | 8.64 | 7.76 | 4.36 | -0.82 | | | 0 | | | M2,25 | |
| 59 | 59 | 33.0 | 0.54 | CC | y | 15.99 | 0.14 | 14.42 | 0.15 | 12.15 | 11.07 | 10.42 | 9.69 | 9.22 | 8.62 | 8.03 | 4.06 | -0.92 | | -3.50 | 3.9 | 0.61 | 3596 | M1,75 | |
| 91 | 92 | 84.7 | 0.32 | CC | y | 16.17 | 0.11 | 14.46 | 0.09 | 12.67 | 11.68 | 11.25 | 10.62 | 10.19 | 9.73 | 8.93 | 6.26 | -0.93 | | -3.82 | 3.19 | 0.33 | 3705 | M1 | 8.4 |
| 93 | 94 | 48.5 | 0.46 | CC | y | 16.98 | 0.16 | 15.36 | 0.23 | 12.89 | 11.87 | 11.34 | 10.04 | 9.59 | 9.14 | 8.42 | 6.16 | -1.00 | y | -3.15 | 4.75 | 0.38 | 3560 | M2 | 0 |
| 96 | 97 | 28.3 | 0.52 | CC | y | 17.93 | 0.14 | 15.76 | 0.05 | 13.17 | 11.93 | 11.44 | 10.77 | 10.61 | 10.26 | 9.39 | 5.1 | -1.27 | y | -3.73 | 5.25 | 0.34 | 3560 | M2 | 19.8 |
| 108 | 110 | 49.3 | 0.31 | CC | y | 16.61 | 0.13 | 14.72 | 0.11 | 12.73 | 11.83 | 11.47 | 11.01 | 10.72 | 10.4 | 9.79 | 7.78 | -1.46 | | -4.82 | 3.55 | 0.32 | 3560 | M2 | 2.2 |
| 130 | 133 | 99.1 | 0.64 | CC | y | 18.65 | 0.12 | 16.66 | 0.33 | 13.88 | 12.58 | 11.58 | 10.28 | 9.7 | 9.26 | 8.62 | 6.01 | -0.97 | | -4.23 | 4.79 | 0.14 | 3451 | M2,75 | 0 |
| 138 | 142 | 45.9 | 0.67 | CC | y | 17.96 | 0.13 | 15.84 | 0.06 | 13.52 | 12.4 | 11.84 | 11.08 | 10.7 | 10.27 | 9.71 | 4.21 | -1.28 | | -4.09 | 3.19 | 0.012 | 3234 | M4,25 | 0 |
| 176 | 186 | 150.0 | 0.47 | CC | y | 19.16 | 0.19 | 17.62 | 0.80 | 15.07 | 13.43 | 12.18 | 10.82 | 10.26 | 9.87 | 9.31 | 6 | -1.16 | | -3.89 | 6.91 | 0.097 | 3778 | M0,5 | 0 |
| 207 | 228 | 74.3 | 0.47 | CC | y | 18.38 | 0.13 | 16.46 | 0.12 | 14.11 | 13.35 | 12.89 | 9.19 | 8.8 | 8.4 | 7.63 | 4.95 | -1.07 | | -3.51 | 1.84 | 0.046 | 3125 | M5 | |
| 223 | 246 | 121.4 | 0.74 | CC | y | 19.69 | 0.46 | 17.64 | 0.17 | 14.86 | 14.02 | 13.5 | 12.84 | 12.39 | 12.09 | 11.39 | 5.86 | -1.24 | | | 2.41 | 0.028 | 3058 | M5,5 | 1.6 |
| 247 | 276 | 105.0 | 0.74 | CC | y | 20.26 | 0.91 | 18.08 | 0.35 | 15.2 | 14.36 | 13.7 | 12.87 | 12.37 | 11.96 | 11.45 | 7.34 | -1.24 | | | 2.91 | 0.022 | 2935 | M6,5 | |
| 251 | 280 | 102.5 | 0.52 | CC | y | 19.81 | 0.44 | 17.53 | 0.47 | 15.38 | 14.61 | 14.2 | 13.71 | 13.47 | 13.44 | | 4.34 | -2.35 | | -2.98 | 1.31 | 0.014 | 3024 | M5,75 | 1.7 |
| 15 | 15 | 42.0 | 0.13 | CC | n | 15.23 | 0.17 | 13.44 | 0.12 | 11.02 | 9.93 | 9.33 | 8.52 | 8.17 | 7.81 | 7.17 | 4.11 | -1.33 | y | -5.11 | 3.97 | 1.9 | 3778 | M0,5 | 0 |
| 56 | 55 | 52.6 | 0.33 | CC | n | 18.31 | | 16.11 | 0.05 | 12.48 | 11.27 | 10.64 | 9.54 | 9.12 | 8.64 | 7.76 | 4.36 | -0.82 | | -3.09 | 3.65 | 0.34 | 3850 | M0 | 6.34 |
| 57 | 57 | 38.9 | 0.27 | CC | n | 16.51 | 0.39 | 15.14 | 0.18 | 12.54 | 11.27 | 10.7 | 10.01 | 9.71 | 9.39 | 8.35 | 4.52 | -0.98 | y | -3.45 | 4.43 | 0.54 | 3955 | K8 | 30 |
| 71 | 72 | 53.8 | 0.30 | CC | n | 17.79 | 0.16 | 15.82 | 0.17 | 12.98 | 11.55 | 10.84 | 9.96 | 9.56 | 9.34 | 8.71 | 5.41 | -1.47 | | -4.15 | 6.42 | 0.53 | 3778 | M0,5 | 0 |
| 116 | 119 | 53.0 | 0.05 | CC | n | 17.76 | | 15.35 | 0.06 | 12.89 | 11.98 | 11.53 | 11.06 | 10.68 | 10.2 | 9.52 | 6.85 | -1.08 | n | -4.42 | 3.05 | 0.2 | 3161 | M4,75 | 0 |
| 127 | 130 | 59.6 | 0.39 | CC | n | 17.95 | 0.21 | 15.72 | 0.20 | 13.21 | 12.22 | 11.75 | 11.03 | 10.67 | 10.38 | 9.74 | 6.97 | -1.41 | | | 4.15 | 0.2 | 3161 | M4,75 | 0 |
| 142 | 146 | 86.3 | 0.19 | CC | n | 18.19 | 0.06 | 15.87 | 0.02 | 13.19 | 12.41 | 11.89 | 11.23 | 10.89 | 10.53 | 9.88 | 6.67 | -1.31 | | | 2.02 | 0.12 | 3024 | M5,75 | 2.2 |
| 159 | 165 | 65.8 | 0.38 | CC | n | 18.74 | 0.21 | 16.19 | 0.19 | 13.74 | 12.66 | 11.95 | 11.28 | 11 | 10.76 | 10.36 | 6.87 | -1.82 | | -4.16 | 3.05 | 0.09 | 3161 | M4,75 | 0 |
| 172 | 182 | 41.0 | 0.27 | CC | n | 18.72 | 0.31 | 16.52 | 0.06 | 14.11 | 13.03 | 12.46 | 11.75 | 11.39 | 11.03 | 9.89 | 2.59 | -0.76 | | -3.65 | 3.23 | 0.068 | 3198 | M4,5 | 0 |
| 181 | 194 | 66.5 | 0.29 | CC | n | | | 16.21 | | 13.84 | 12.83 | 12.39 | 11.9 | 11.61 | 11.34 | 10.78 | 8.85 | -1.59 | | -3.86 | 3.19 | 0.086 | 3198 | M4,5 | |
| 190 | 204 | 46.0 | 0.38 | CC | n | 18.35 | 0.20 | 16.08 | 0.15 | 13.61 | 12.99 | 12.56 | 11.95 | 11.55 | 11.13 | 10.33 | 7.76 | -1.01 | | | 1.28 | 0.068 | 3024 | M5,75 | |
| | 113 | 31.7 | 0.18 | CC | n | 17.18 | 0.34 | 15.39 | 0.33 | 12.95 | 11.98 | 11.46 | | | | | | | | | | 0.042 | 3342 | K6.5 | |
| 20 | 20 | 3.4 | 0.55 | CW | y | 14.72 | 0.31 | 13.20 | 0.44 | 11.02 | 9.99 | 9.47 | 8.99 | 8.8 | 8.71 | 8.15 | 3.57 | -1.94 | y | -5.59 | 5.39 | 3.9 | 5250 | K0 | |
| 24 | 24 | 20.7 | 0.65 | CW | y | 18.16 | 0.07 | 16.51 | 0.13 | 12.08 | 10.94 | 10.39 | 9.7 | 9.48 | 8.92 | 8.25 | 5.44 | -1.14 | y | -3.52 | 3.79 | 0.69 | 3850 | M0 | |
| 34 | 34 | 18.9 | 0.81 | CW | y | 15.01 | 0.37 | 13.52 | 0.32 | 11.45 | 10.44 | 9.87 | 9.19 | 8.8 | 8.4 | 7.63 | 4.95 | -1.07 | y | -3.49 | 2.8 | 0.99 | 4205 | K6 | 8.6 |
| 53 | 53 | 17.9 | 0.83 | CW | y | 15.86 | 0.12 | 14.13 | 0.09 | 11.94 | 10.9 | 10.47 | 9.78 | 9.56 | 9.24 | 8.46 | 4.62 | -1.35 | y | -3.12 | 3.69 | 0.72 | 3669 | M1,25 | 7.3 |
| 81 | 82 | 11.1 | 0.87 | CW | y | 16.49 | 0.15 | 14.72 | 0.11 | 12.59 | 11.52 | 11.14 | 10.65 | 10.32 | 9.96 | 9.25 | 2.64 | -1.26 | y | -3.12 | 3.83 | 0.39 | 3560 | M2 | 3.9 |
| 102 | 104 | 13.8 | 0.78 | CW | y | 16.80 | 0.17 | 15.01 | 0.17 | 12.8 | 11.84 | 11.46 | 11.15 | 11.05 | 11.04 | 11.15 | 6.34 | -2.88 | | -2.80 | 3.97 | 0.32 | 3488 | M2,5 | 6.9 |
| 111 | 114 | 13.9 | 0.60 | CW | y | 18.42 | 0.12 | 16.09 | 0.14 | 13.3 | 12.05 | 11.49 | 10.96 | 10.84 | 10.62 | 10.07 | 5.39 | -1.84 | | -3.59 | 3.9 | 0.17 | 3125 | M5 | 2.1 |
| 124 | 127 | 14.7 | 0.76 | CW | y | 17.67 | 0.27 | 15.64 | 0.20 | 13.07 | 12.1 | 11.66 | 11.01 | 10.79 | 10.51 | 9.94 | 4.07 | -1.64 | | -4.07 | 3.44 | 0.18 | 3161 | M4,75 | 2.5 |
| 129 | 132 | 30.0 | 0.48 | CW | y | 17.35 | 0.30 | 15.31 | 0.22 | 13 | 12.12 | 11.76 | 11 | 10.74 | 10.45 | 9.75 | 6.77 | -1.44 | | -4.06 | 2.7 | 0.17 | 3234 | M4,25 | 1.3 |
| 135 | 139 | 35.0 | 0.93 | CW | y | 18.48 | 0.27 | 15.94 | 0.22 | 13.28 | 12.33 | 11.79 | 11.22 | 10.93 | 10.49 | 9.64 | 2.62 | -1.04 | n | -4.71 | 3.62 | 0.16 | 3091 | M5,25 | 1.9 |
| 147 | 151 | 7.7 | 0.60 | CW | y | 16.90 | 0.35 | 15.23 | 0.35 | 13.11 | 12.22 | 11.96 | 11.75 | 11.73 | 11.63 | 11.44 | 6.55 | -2.50 | n | -3.25 | 1.81 | 0.13 | 3451 | M2,75 | 6.9 |
| 158 | 164 | 26.7 | 0.91 | CW | y | 17.61 | 0.15 | 15.50 | 0.07 | 13.5 | 12.52 | 12.02 | 11.5 | 11.25 | 10.86 | 10.19 | 4.79 | -1.34 | | -4.92 | 3.05 | 0.12 | 3270 | M4 | 0 |
| 163 | 171 | 21.1 | 0.62 | CW | y | 19.07 | 1.10 | 17.41 | 0.98 | 16.04 | 13.9 | 12.27 | 10.38 | 9.87 | 9.57 | 8.94 | 6.03 | -1.25 | y | -1.45 | 4.54 | 0.021 | 3741 | M0,75 | 0 |
| 167 | 176 | 17.7 | 0.68 | CW | y | 19.65 | 1.90 | 17.34 | 1.47 | 13.69 | 12.86 | 12.48 | 11.91 | 11.62 | 11.42 | 10.95 | 8.52 | -1.79 | | | 2.09 | 0.073 | 3161 | M4,75 | 2.3 |
| 179 | 192 | 10.0 | 1.02 | CW | y | 17.77 | 0.16 | 15.62 | 0.10 | 13.56 | 12.76 | 12.36 | 11.97 | 11.7 | 11.41 | 10.93 | 6.61 | -1.68 | | -3.88 | 1.7 | 0.074 | 3125 | M5 | 1.7 |
| 202 | 222 | 25.3 | 0.67 | CW | y | 19.49 | 0.37 | 16.95 | 0.18 | 14.07 | 13.28 | 12.73 | 11.97 | 11.56 | 11.15 | 10.39 | 7.84 | -1.07 | | -4.71 | 1.06 | 0.032 | 2838 | M7,25 | 3.4 |
| 238 | 264 | 22.2 | 0.37 | CW | y | 19.20 | 0.28 | 17.06 | 0.10 | 14.84 | 14.14 | 13.74 | 13.37 | 13.17 | 13.29 | 12.98 | 5.9 | -2.49 | | -3.49 | 1.24 | 0.021 | 3058 | M5,5 | 0 |
| 271 | 304 | 20.2 | 0.60 | CW | y | 17.75 | 0.02 | 15.77 | 0.25 | 15.66 | 15.33 | 14.54 | 14.73 | 14.38 | | | 7.39 | -1.53 | | -2.91 | 1.74 | 0.009 | 2880 | M7 | |
| M185 | | 38.9 | 0.57 | CW | y | 17.18 | 0.17 | 15.25 | 0.12 | 13 | 12.23 | 11.91 | | | | | | | | | 1.5 | 0.134 | 3320 | M5,5 | |
| | 103 | 20.7 | 0.65 | CW | y | 17.24 | 0.22 | 15.94 | 0.35 | 13.49 | 12.29 | 11.39 | | | | | | | | -4.17 | | | 3342 | M4 | 17.7 |
| 31 | 31 | 2.7 | 0.85 | WW | y | 14.50 | 0.53 | 13.10 | 0.26 | 11.29 | 10.38 | 10.07 | 9.86 | 9.93 | 9.76 | 9.54 | 3.96 | -2.46 | n | -4.30 | 2.52 | 0.87 | 3488 | M2,5 | 0 |
| 33 | 33 | 3.0 | 0.51 | WW | y | 14.50 | 0.09 | 13.09 | 0.10 | 11.18 | 10.23 | 9.85 | 9.46 | 9.2 | 8.58 | 7.55 | 4.74 | -0.62 | y | -5.35 | 3.37 | 1.5 | 4205 | K6 | 5.1 |
| 47 | 47 | 1.3 | 0.43 | WW | y | 16.14 | 0.16 | 14.33 | 0.25 | 11.85 | 10.75 | 10.36 | 10.07 | 10.07 | 9.96 | 9.97 | 3.73 | -2.73 | n | -3.52 | 5.39 | 1.4 | 4590 | K4 | 3.1 |
| 49 | 49 | 2.2 | 0.41 | WW | y | 16.74 | 0.07 | 14.82 | 0.12 | 12.12 | 10.89 | 10.42 | 10.08 | 10.07 | 9.91 | 9.78 | 4.55 | -2.49 | n | -3.70 | 5.25 | 0.92 | 3705 | M1 | 8 |
| 61 | 61 | 1.7 | 0.49 | WW | y | 14.80 | 0.13 | 13.54 | 0.13 | 11.85 | 10.98 | 10.7 | 10.61 | 10.58 | 10.49 | 10.34 | 7.21 | -2.56 | y | -3.39 | 2.16 | 0.56 | 3850 | M0 | 16.4 |
| 66 | 67 | 3.0 | 0.77 | WW | y | 16.20 | 0.12 | 14.02 | 0.44 | 12.11 | 11.13 | 10.76 | 10.45 | 10.34 | 10.1 | 9.74 | 4.5 | -2.03 | y | -3.65 | 3.37 | 0.47 | 3415 | M3 | 6.7 |
| 68 | 69 | 3.7 | 0.50 | WW | y | 16.33 | 0.01 | 14.34 | 0.03 | 12.14 | 11.13 | 10.81 | 10.54 | 10.46 | 10.4 | 10.52 | 4.1 | -2.84 | n | -3.95 | 4.04 | 0.62 | 3560 | M2 | |
| 75 | 76 | 4.4 | 0.48 | WW | y | 16.41 | 0.13 | 14.79 | 0.17 | 12.49 | 11.44 | 10.99 | 10.33 | 10.07 | 9.78 | 8.85 | 2.54 | -1.18 | y | -3.54 | 4.29 | 0.51 | 3705 | M1 | 8.4 |
| 76 | 77 | 4.0 | 0.35 | WW | y | 16.40 | 0.08 | 14.46 | 0.35 | 12.23 | 11.3 | 11.05 | 10.78 | 10.77 | 10.65 | 10.69 | 6.29 | -2.75 | | -3.66 | 2.13 | 0.3 | 3379 | M3,25 | 2.7 |
| 77 | 78 | 3.3 | 0.36 | WW | y | 15.71 | 0.08 | 14.00 | 0.10 | 12.13 | 11.27 | 10.97 | 10.7 | 10.73 | 10.68 | 10.79 | 5.93 | -2.96 | y | -3.94 | 2.8 | 0.46 | 3560 | M2 | 5.4 |
| 78 | 79 | 6.2 | 0.66 | WW | y | 15.67 | 0.01 | 14.28 | 0.27 | 12.24 | 11.3 | 11.06 | 10.82 | 10.86 | 10.76 | 10.72 | 6.93 | -2.72 | | -2.19 | 1.67 | 0.33 | 3741 | M0,75 | 14 |
| 79 | 80 | 3.8 | 0.43 | WW | y | 16.41 | 0.18 | 14.56 | 0.07 | 12.48 | 11.47 | 11.1 | 10.75 | 10.68 | 10.47 | 10.29 | 4.45 | -2.31 | n | -3.31 | 3.65 | 0.36 | 3379 | M3,25 | 5.5 |
| 80 | 81 | 2.0 | 0.97 | WW | y | 16.14 | 0.24 | 14.55 | 0.00 | 12.54 | 11.47 | 11.01 | 10.31 | 10.09 | 9.87 | 9.77 | 2.51 | -2.24 | y | -3.68 | 3.83 | 0.41 | 3560 | M2 | 2.2 |
| 82 | 83 | 4.5 | 0.54 | WW | y | 15.85 | 0.07 | 14.10 | 0.11 | 12.24 | 11.37 | 11.08 | 10.87 | 10.81 | 10.76 | 10.76 | 5.94 | -2.74 | | -3.51 | 2.87 | 0.39 | 3488 | M2,5 | 9.7 |

**Table 4 – continued from previous page**

| N1 (1) | N2 (2) | EW (3) | $f_v$ (4) | Class. (5) | $V_{EW}$ (6) | <$R_c$> (7) | sd $R_c$ (8) | <$I_c$> (9) | sd $I_c$ (10) | J (11) | H (12) | K (13) | [3.6] (14) | [4.5] (15) | [5.8] (16) | [8.0] (17) | [24] (18) | $\alpha$ (19) | $V_{MIR}$ (20) | $\mathrm{Log}(L_x/L_{bol})$ (21) | $A_v$ (22) | Lum (23) | $T_{ef}$ (24) | Sp (25) | P(d) (26) |
|---|---|---|---|---|---|---|---|---|---|---|---|---|---|---|---|---|---|---|---|---|---|---|---|---|---|
| 83 | 84 | 5.7 | 0.44 | WW | y | 15.57 | 0.14 | 13.93 | 0.10 | 12.18 | 11.34 | 11.1 | 10.86 | 10.82 | 10.77 | 10.8 | 6.76 | -2.80 | | -2.86 | 1.77 | 0.31 | 3488 | M2,5 | 4.5 |
| 85 | 86 | 6.3 | 0.37 | WW | y | 16.26 | 0.09 | 14.29 | 0.07 | 12.31 | 11.42 | 11.12 | 10.84 | 10.77 | 10.7 | 10.64 | 5.86 | -2.63 | | -3.88 | 2.34 | 0.29 | 3270 | M4 | 14 |
| 87 | 88 | 4.1 | 0.92 | WW | y | 17.92 | 0.28 | 15.71 | 0.33 | 11.82 | 10.95 | 10.68 | 10.64 | 10.42 | 10.23 | 9.55 | 5.98 | -1.63 | y | -3.75 | 6.03 | 0.54 | 3524 | M0,5 | 7.3 |
| 88 | 89 | 5.6 | 0.42 | WW | y | 16.92 | 0.05 | 14.81 | 0.07 | 12.47 | 11.52 | 11.15 | 10.94 | 10.74 | 10.74 | 10.82 | 7.07 | -2.76 | n | -3.55 | 3.3 | 0.33 | 3270 | M4 | 0 |
| 89 | 90 | 6.9 | 0.35 | WW | y | 16.72 | 0.30 | 14.60 | 0.15 | 12.89 | 11.9 | 11.35 | 10.72 | 10.47 | 9.94 | 9.25 | 7.19 | -1.13 | | -3.30 | 2.62 | 0.26 | 3306 | M3,75 | 0 |
| 94 | 95 | 2.9 | 0.43 | WW | y | 15.78 | 0.33 | 14.29 | 0.24 | 12.39 | 11.47 | 11.2 | 10.95 | 10.97 | 10.94 | 11.01 | 5.67 | -2.93 | y | -3.43 | 2.7 | 0.39 | 3850 | M0 | 9.1 |
| 95 | 96 | 2.5 | 0.52 | WW | y | 16.11 | 0.07 | 14.34 | 0.08 | 12.49 | 11.57 | 11.27 | 11.01 | 10.93 | 10.94 | 11.32 | 3.48 | -3.23 | n | -3.68 | 2.13 | 0.24 | 3379 | M3,25 | 2.1 |
| 99 | 100 | 2.4 | 0.73 | WW | y | 15.56 | 0.09 | 14.25 | 0.10 | 12.63 | 11.64 | 11.38 | 11.21 | 11.17 | 10.98 | 11.07 | 3.1 | -2.68 | y | -3.88 | 2.73 | 0.32 | 4205 | K6 | 0 |
| 101 | 102 | 4.0 | 0.62 | WW | y | 16.20 | 0.16 | 14.56 | 0.08 | 12.66 | 11.7 | 11.4 | 11.26 | 11.04 | 10.96 | 10.93 | 6.05 | -2.52 | y | -2.55 | 2.87 | 0.29 | 3632 | M1,5 | 7 |
| 103 | 105 | 3.0 | 0.46 | WW | y | 16.06 | 0.10 | 14.37 | 0.06 | 12.61 | 11.66 | 11.4 | 11.12 | 11.12 | 11.12 | 11.13 | 6.35 | -2.89 | | -4.12 | 1.88 | 0.22 | 3524 | M2,25 | 5.1 |
| 105 | | 4.3 | 0.42 | WW | y | 17.12 | 0.16 | 15.31 | 0.14 | 12.85 | 11.81 | 11.42 | 11.21 | 11.03 | 10.99 | 11.15 | 6.35 | -2.82 | | | | | 3705 | M1 | |
| 107 | 109 | 3.0 | 0.48 | WW | y | 16.20 | 0.14 | 14.50 | 0.14 | 12.52 | 11.59 | 11.3 | 11.06 | 11.01 | 11.04 | 10.91 | 5.83 | -2.71 | y | -3.36 | 2.77 | 0.29 | 3451 | M2,75 | 8 |
| 109 | 111 | 2.3 | 0.49 | WW | y | 16.40 | 0.06 | 14.60 | 0.08 | 12.7 | 11.69 | 11.37 | 10.68 | 10.48 | 10.26 | 9.29 | 6.67 | -1.29 | | -3.63 | 2.8 | 0.27 | 3560 | M2 | 0 |
| 110 | 112 | 12.5 | 0.54 | WW | y | 17.26 | 0.08 | 15.05 | 0.20 | 12.58 | 11.76 | 11.43 | 11.07 | 10.98 | 10.9 | 10.91 | 5.06 | -2.68 | | -3.96 | 2.34 | 0.22 | 3161 | M4,75 | 1.2 |
| 118 | 121 | 3.1 | 0.54 | WW | y | 16.36 | 0.42 | 14.81 | 0.27 | 12.82 | 11.82 | 11.48 | 11.29 | 11.25 | 11.19 | 10.95 | 4.79 | -2.49 | n | -4.28 | 3.87 | 0.35 | 3778 | M0,5 | 0 |
| 120 | 123 | 2.4 | 0.38 | WW | y | 16.06 | 0.15 | 14.63 | 0.14 | 12.91 | 11.92 | 11.69 | 11.48 | 11.46 | 11.37 | 11.4 | 6.65 | -2.76 | | -3.66 | 2.2 | 0.21 | 3850 | M0 | 13.5 |
| 121 | 124 | 5.5 | 0.32 | WW | y | 16.78 | 0.10 | 14.66 | 0.07 | 12.65 | 11.8 | 11.47 | 11.18 | 11.01 | 11 | 11.16 | 3.99 | -2.87 | | -3.79 | 1.7 | 0.17 | 3161 | M4,75 | 3.7 |
| 122 | 125 | 2.3 | 0.58 | WW | y | 15.12 | 0.08 | 13.93 | 0.07 | 12.55 | 11.74 | 11.55 | 11.44 | 11.38 | 11.26 | 12.6 | 4.78 | -4.13 | n | -3.38 | 0.71 | 0.19 | 3705 | M1 | 12 |
| 128 | 131 | 7.6 | 0.47 | WW | y | 17.63 | 0.32 | 15.24 | 0.24 | 13.01 | 12.09 | 11.7 | 11.36 | 11.23 | 11.15 | 11.23 | 4.96 | -2.71 | n | -3.44 | 3.19 | 0.19 | 3198 | M4,5 | 3.3 |
| 134 | 138 | 10.5 | 0.47 | WW | y | 17.32 | 0.22 | 15.02 | 0.20 | 12.78 | 12.07 | 11.73 | 11.32 | 11.28 | 11.29 | 11.22 | 7.1 | -2.77 | | -4.12 | 1.42 | 0.14 | 3091 | M5,25 | 13 |
| 141 | 145 | 6.7 | 0.56 | WW | y | 17.27 | 0.12 | 15.30 | 0.12 | 13.19 | 12.17 | 11.84 | 11.52 | 11.47 | 11.24 | 11.84 | 7.37 | -3.17 | n | -3.55 | 3.16 | 0.17 | 3451 | M2,75 | 0 |
| 148 | 153 | 2.6 | 0.53 | WW | y | 16.86 | 0.14 | 14.99 | 0.09 | 13.13 | 12.29 | 12.01 | 11.81 | 11.68 | 11.66 | 11.61 | 7.4 | -2.66 | | -3.73 | 1.84 | 0.12 | 3342 | M3,5 | 0 |
| 160 | 166 | 4.7 | 0.65 | WW | y | 18.27 | 0.35 | 15.85 | 0.18 | 13.38 | 12.54 | 12.09 | 11.73 | 11.59 | 11.55 | 11.43 | 6.74 | -2.54 | | -3.77 | 2.55 | 0.12 | 3058 | M5,5 | 0 |
| 166 | 174 | 7.5 | 0.51 | WW | y | 17.67 | 0.10 | 15.70 | 0.11 | 13.52 | 12.59 | 12.22 | 14.73 | 14.38 | | | 7.39 | -1.53 | | -3.35 | 3.19 | 0.12 | 3342 | M3,5 | 8.6 |
| 169 | 179 | 6.1 | 0.40 | WW | y | 18.39 | 0.41 | 16.03 | 0.13 | 13.69 | 12.7 | 12.35 | 11.97 | 11.89 | 11.84 | 11.88 | 7.17 | -2.76 | | -3.50 | 3.44 | 0.11 | 3270 | M4 | |
| 173 | 183 | 8.5 | 0.92 | WW | y | 17.99 | 0.22 | 15.73 | 0.08 | 13.56 | 12.77 | 12.44 | 12.08 | 12.01 | 11.94 | 12.09 | 6.76 | -2.87 | | -3.40 | 1.7 | 0.074 | 3125 | M5 | |
| 186 | 200 | 8.5 | 0.61 | WW | y | 17.63 | 0.16 | 15.62 | 0.14 | 13.7 | 12.88 | 12.55 | 12.33 | 12.18 | 12.21 | 12.1 | 4.81 | -2.64 | | -2.60 | 1.95 | 0.072 | 3198 | M4,5 | 0 |
| 195 | 211 | 8.2 | 0.54 | WW | y | 18.05 | 0.17 | 15.94 | 0.11 | 13.73 | 12.93 | 12.57 | 12.19 | 12.04 | 11.91 | 11.38 | 0.34 | -1.96 | | -3.90 | 2.45 | 0.078 | 3161 | M4,75 | 2.2 |
| 209 | 230 | 4.2 | 0.79 | WW | y | 19.33 | 0.13 | 16.89 | 0.16 | 14.24 | 13.32 | 12.85 | 12.46 | 12.34 | 12.73 | 12.86 | 5.11 | -3.43 | | -3.62 | 4.15 | 0.075 | 3161 | M4,75 | 0 |
| 233 | 256 | 5.9 | 0.60 | WW | y | 18.55 | 0.06 | 16.43 | 0.08 | 14.54 | 13.84 | 13.54 | 13.26 | 13.13 | 12.96 | | | -2.30 | | -3.47 | 0.85 | 0.024 | 3161 | M4,75 | 2.3 |
| 45 | 45 | 2.1 | 0.21 | WW | n | 14.51 | 0.10 | 13.26 | 0.12 | 11.5 | 10.6 | 10.31 | 10.14 | 10.06 | 10.02 | 10.02 | 5.22 | -2.74 | y | -3.52 | 3.01 | 0.99 | 4278 | K5,5 | 5.4 |
| 58 | 58 | 10.9 | 0.08 | WW | n | 16.22 | 0.12 | 14.24 | 0.25 | 11.8 | 10.94 | 10.59 | 10.44 | 10.34 | 10.29 | 10.41 | 7.34 | -2.84 | y | -3.00 | 1.84 | 0.39 | 3161 | M4,75 | 3.1 |
| 60 | 60 | 2.5 | 0.17 | WW | n | 15.06 | 0.03 | 13.61 | 0.14 | 11.82 | 10.95 | 10.68 | 10.49 | 10.57 | 10.44 | 10.38 | 7.39 | -2.71 | | -2.93 | 1.28 | 0.45 | 3778 | M0.5 | 8.4 |
| 65 | 65 | 3.0 | 0.25 | WW | n | 15.01 | 0.20 | 13.65 | 0.15 | 11.95 | 11.14 | 10.85 | 10.69 | 10.67 | 10.59 | 10.68 | 4.61 | -2.85 | y | -3.30 | 1.95 | 0.46 | 3705 | M1 | 9.1 |
| 98 | 99 | 9.2 | 0.21 | WW | n | 17.42 | 0.13 | 15.03 | 0.09 | 12.6 | 11.7 | 11.33 | 10.91 | 10.87 | 10.83 | 10.8 | 6.49 | -2.74 | | -2.25 | 2.7 | 0.24 | 3161 | M4,75 | |
| 119 | 122 | 3.3 | 0.17 | WW | n | 16.05 | 0.20 | 14.60 | 0.15 | 12.63 | 11.73 | 11.48 | 11.29 | 11.23 | 11.18 | 11.2 | 6.16 | -2.77 | n | -3.43 | 2.94 | 0.33 | 3850 | M0 | |
| 139 | 143 | 6.1 | 0.17 | WW | n | 18.03 | 0.20 | 15.69 | 0.15 | 13.15 | 12.28 | 11.84 | 11.44 | 11.37 | 11.29 | 11.03 | 4.28 | -2.40 | | -4.12 | 2.77 | 0.15 | 3091 | M5,25 | 0 |
| 143 | 147 | 3.8 | 0.19 | WW | n | 16.49 | 0.11 | 15.06 | 0.33 | 13.02 | 12.13 | 11.84 | 11.61 | 11.58 | 11.55 | 11.47 | 7.45 | -2.71 | n | -3.12 | 2.27 | 0.18 | 3632 | M1,5 | |
| 150 | 156 | 6.5 | 0.33 | WW | n | 18.14 | 0.21 | 15.86 | 0.19 | 13.3 | 12.31 | 11.91 | 11.59 | 11.49 | 11.38 | 11.47 | 7.47 | -2.72 | n | -4.02 | 3.9 | 0.18 | 3270 | M4 | 9.6 |
| 152 | 158 | 4.8 | 0.22 | WW | n | 17.95 | 0.58 | 16.04 | 0.28 | 13.31 | 12.42 | 12 | 11.68 | 11.56 | 11.48 | 11.57 | 6.83 | -2.74 | | -3.72 | 4.65 | 0.15 | 3234 | M4,25 | |
| 154 | 160 | 4.5 | 0.16 | WW | n | 16.93 | 0.52 | 14.89 | 0.05 | 13.08 | 12.23 | 11.98 | 11.76 | 11.72 | 11.68 | 11.59 | 6.79 | -2.69 | n | -3.72 | 1.81 | 0.13 | 3451 | M2,75 | 3.3 |
| 162 | 169 | 4.0 | 0.35 | WW | n | 17.47 | | 15.35 | 0.14 | 13.28 | 12.4 | 12.12 | 11.81 | 11.76 | 11.72 | 11.64 | 7.43 | -2.68 | | -4.03 | 1.95 | 0.11 | 3270 | M4 | |
| 168 | 177 | 9.5 | 0.39 | WW | n | 17.75 | 0.14 | 15.60 | 0.03 | 13.36 | 12.6 | 12.22 | 11.82 | 11.65 | 11.3 | 10.73 | 7.44 | -1.59 | | -3.80 | 2.09 | 0.099 | 3161 | M4,75 | 0 |
| 170 | 180 | 8.1 | 0.33 | WW | n | 18.35 | 0.09 | 16.03 | 0.08 | 13.54 | 12.64 | 12.21 | 11.91 | 11.72 | 11.55 | 11.39 | 5.03 | -2.27 | | -4.26 | 3.19 | 0.11 | 3125 | M5 | 0 |
| 171 | 181 | 9.5 | 0.22 | WW | n | 19.11 | 0.35 | 16.68 | 0.12 | 13.8 | 12.79 | 12.27 | 11.83 | 11.71 | 11.52 | 11.29 | 7.29 | -2.23 | | -3.96 | 4.08 | 0.11 | 3091 | M5,25 | 0 |
| 175 | 185 | 3.2 | 0.33 | WW | n | 18.11 | 0.18 | 15.77 | 0.12 | 13.51 | 12.75 | 12.36 | 11.93 | 11.86 | 11.87 | 11.33 | 4.27 | -2.22 | n | -3.58 | 1.67 | 0.078 | 3091 | M5,25 | 1.4 |
| 187 | 201 | 9.6 | 0.12 | WW | n | 18.44 | 0.18 | 15.99 | 0.09 | 13.58 | 12.82 | 12.42 | 12.04 | 11.93 | 11.97 | 12.22 | 7.43 | -3.10 | | -4.00 | 2.2 | 0.086 | 3058 | M5,5 | 0 |
| 189 | 203 | 8.8 | 0.27 | WW | n | 18.25 | 0.15 | 15.99 | 0.12 | 13.7 | 13.01 | 12.62 | 12.25 | 12.2 | 11.97 | 11.83 | 4.95 | -2.35 | | -3.49 | 0.78 | 0.056 | 3024 | M5,75 | 0 |
| 197 | 216 | 4.0 | 0.35 | WW | n | 18.12 | 0.19 | 16.05 | 0.08 | 13.91 | 13.1 | 12.75 | 12.41 | 12.29 | 12.26 | 12.99 | 3.75 | -3.52 | | -4.16 | 1.95 | 0.057 | 3125 | M5 | 1.5 |
| 198 | 217 | 6.3 | 0.06 | WW | n | 19.05 | 0.33 | 16.52 | 0.32 | 14.03 | 13.18 | 12.76 | 12.44 | 12.28 | 12.37 | | 6.6 | -2.73 | | -4.29 | 2.45 | 0.061 | 3058 | M5,5 | 0 |
| 206 | 227 | 11.0 | 0.13 | WW | n | 18.81 | 0.22 | 16.44 | 0.11 | 13.98 | 13.26 | 12.82 | 12.38 | 12.28 | 12.24 | | 4.75 | -2.59 | | -4.05 | 1.1 | 0.047 | 2990 | M6 | 0 |
| 212 | 233 | 10.0 | 0.00 | WW | n | 19.15 | 0.08 | 16.54 | 0.21 | 14.28 | 13.5 | 13.09 | 12.66 | 12.52 | 12.45 | 13.41 | 4.93 | -3.69 | | -4.47 | 6.38 | 0.14 | 3415 | M3 | |
| 227 | 250 | 6.6 | 0.28 | WW | n | 18.66 | 0.18 | 16.56 | 0.13 | 14.46 | 13.65 | 13.29 | 12.95 | 12.8 | 12.73 | 12.3 | 6.07 | -2.15 | | -3.64 | 2.3 | 0.038 | 3161 | M4,75 | 1.5 |
| 228 | 251 | 9.5 | 0.07 | WW | n | 19.25 | 0.19 | 16.88 | 0.19 | 14.32 | 13.6 | 13.22 | 12.82 | 12.74 | 12.61 | 12.65 | 6.26 | -2.65 | | -3.98 | 1.56 | 0.038 | 3024 | M5,75 | 0 |
| 232 | 255 | 7.3 | 0.26 | WW | n | 19.03 | 0.19 | 16.82 | 0.11 | 14.61 | 13.92 | 13.56 | 13.22 | 13.07 | 13.04 | 12.98 | 7.25 | -2.62 | | -3.57 | 0.92 | 0.024 | 3058 | M5,5 | |
| 23 | 23 | 0.5 | | WAbs | | 14.05 | | 12.71 | | 10.99 | 10.07 | 9.76 | 9.51 | 9.49 | 9.34 | 8.89 | 4.13 | -2.15 | | -5.63 | 2.84 | 1.5 | 4132 | K6,5 | 4.5 |
| 32 | 32 | -0.5 | | WAbs | | 14.60 | 0.15 | 13.10 | 0.14 | 10.84 | 9.96 | 9.56 | 9.37 | 9.33 | 9.17 | 9.15 | 5.31 | -2.59 | n | -4.38 | 5.07 | 3.6 | 4730 | K3 | 5.2 |
| 52 | 52 | 0.5 | | WAbs | | 14.26 | 0.34 | 13.11 | 0.27 | 11.55 | 10.71 | 10.5 | 10.41 | 10.4 | 10.31 | 10.34 | 6.93 | -2.77 | n | -3.98 | 2.62 | 0.26 | 3609 | K3,5 | |
| 62 | 62 | 1 | | WAbs | | 14.51 | 0.11 | 13.32 | 0.15 | 11.67 | 10.85 | 10.58 | 10.52 | 10.51 | 10.6 | 10.51 | 7.29 | -2.89 | y | -3.63 | 2.3 | 0.71 | 4132 | K6,5 | 7 |
| 126 | 129 | 2 | | WAbs | | 16.77 | 0.18 | 14.82 | 0.04 | 12.87 | 11.96 | 11.62 | 11.28 | 11.2 | 11.2 | 11.15 | 7.35 | -2.73 | n | -4.02 | 3.3 | 0.26 | 3560 | M2 | 0 |
| 149 | 155 | 6 | | WAbs | | 18.21 | 0.28 | 15.71 | 0.26 | 13.21 | 12.3 | 11.87 | 11.45 | 11.24 | 11.02 | 10.56 | 7.37 | -1.85 | | -4.33 | 3.05 | 0.15 | 3234 | M4,25 | 2.7 |

According to these data, the following conclusions can be drawn. The first is that a percentage of the number of periodic stars among WTT of objects (66%) is higher, than among CTT (46%). The objects from the intermediate sample CW, as well from the sample WAbs, have the largest percentage of periodic variables (81%). In both samples (CC and WW), among the variable EW($H_\alpha$) objects the percentage of objects with periodical brightness variability is higher.

We did not find a correlation between the rotation period and the $H_\alpha$ activity (EW($H_\alpha$)). This is in agreement with the findings of Cieza & Baliber (2006), where the evolutionary stage of stellar objects was determined based on their IR excess.

We would like to draw attention to one more fact. The averaged values of extinction ($A_v$ in Table 3) in the samples decrease with decreasing H$\alpha$ activity, which, apparently, could be explained by a decrease in the mass and density of circumstellar matter. In addition, in the CC and WW samples the same trend is observed: the average values of $A_v$ are higher for the variables.

### *3.2.2 Infrared range*

One of the main distinguishing properties of young stellar objects is their infrared excess, which is also used for the classification of their evolutionary stage. We determined the average colour indices of the near- and mid-infrared ranges for all samples. The observed and corrected for extinction infrared colour indices are given in Table 3. In addition, we considered the position of the stellar objects in the different samples in the near- and mid-infrared two-colour diagrams (hereafter NIR and MIR c-c diagrams). The diagrams are presented in Fig. 4.

The averaged values of both observed J-K, [3.6]-[8.0] and corrected for extinction $(J\text{-}K)_0$, $([3.6]\text{-}[8.0])_0$ colour indices show that, with a decreasing of the EW($H_\alpha$), the infrared excess is also decreases from CC to WW sample. It should be noted that concerning the values of J-K colours, there is no difference between CC and CW samples. In this regard, we would draw attention to one fact. The CW sample includes the stellar object (N1 = 163, Luhman et al., 2003) with the largest colour index J-K = 3.77 (or corrected for extinction $(J\text{-}K)_0$ = 3.0). Concerning the J-K colour, as well as the position on the JH/HK diagram (see Fig. 4) this star can be considered as an object with Class I evolutionary stage (Lada & Lada, 2003), though the $\alpha^{3-8\mu m}$ is equal to -1.25 that correspond to Class II evolutionary stage (see text below). If we exclude this object from CW sample, then the average $(J\text{-}K)_0$ decreases (0.95 ± 0.31) and will be intermediate between CC and WW samples. There is no noticeable difference in near-infrared colour indices for variable and non-variable objects in both CC and WW samples. In mid-infrared range there is a slight difference in the WW sample.

The position of objects in the NIR and MIR c-c diagrams also show a well-defined relationship between the equivalent width and the infrared excess. In NIR c-c diagram, objects from the CC sample, with greater EW($H_\alpha$), are located noticeably higher and to the right of the WW stars, which are concentrated between two reddening vectors in the vicinity of dwarf and giant loci. The objects from the CW sample occupy an intermediate position. Similarly, in the MIR c-c diagram the vast majority of the stars from the CC and CW samples have the [3.6]-[8.0] colour index greater than 0.75. Conversely, the colour index of vast majority of objects from the WW sample with a smaller equivalent width of $H_\alpha$ does not exceed 0.5.

Another parameter ($\alpha$ slope), which is used to characterize the shape of the infrared excess in the spectral energy distribution (SED), is defined as $\lambda F_\lambda \propto \lambda^\alpha$. The values of $\alpha$ are from Lada et al. (2006), where it was measured over the 3-8 $\mu$m IRAC bands. The values range from $\alpha$ < -2.56 for purely photospheric emission (Class III), -2.56 < $\alpha$ < -1.8 for evolved disks (Class II/III), and -1.8 < $\alpha$ < 0 for full disks (Class II). The distribution of the $\alpha$ slopes of the SEDs in our samples also reflects a correlation between the infrared excess and $H_\alpha$ activity (see Table 3). According to the $\alpha$ slope, there are only two objects in the CC sample ($\backsim$ 6% of the total number), which cannot be classified as stellar objects with II evolutionary class, including in one case, a star with $\alpha$ = -1.81, which is the boundary value. In the WW sample, only 7 stars ($\backsim$ 10% of the total number) can be classified as objects with II evolutionary class on the basis of their $\alpha$ slope. We would like to note that, according to the results of similar statistics without taking into account the variability of EW($H_\alpha$), the discrepancy between the $\alpha$ slope and $H_\alpha$ activity was somewhat higher: $\backsim$ 10% and $\backsim$ 22% for the CTT and WTT objects, respectively (Nikoghosyan et al., 2015). Similarly, according to the diagram in the Fig. 1 in Stelzer et al. (2012), the discrepancy between $H_\alpha$ activity and the $\alpha^{3-8\,\mu m}$ slopes is higher. In that paper, the variability of EW($H_\alpha$) was also not taken into account.

Among the non-variables of the WW sample, the percentage of objects with $\alpha$ < -2.56 is slightly higher. In the sample CW one can see a well-defined relationship between the values of $\alpha$ and EW($H_\alpha$). This is reflected on the Fig. 5, where the distribution of EW($H_\alpha$) versus $\alpha$ is presented. In CC and WW samples such a relationship is not observed.

In Flaherty et al. (2013) the results of studying the variability of the stellar objects in IC 348 in the 3.6 $\mu$m and 4.5 $\mu$m bands are presented. Unfortunately, only 51 objects from our sample had good data. Nevertheless, although the information is not complete, it confirms the conclusions of Flaherty et al. (2013), that the percentage of MIR variables increases with increasing $H_\alpha$ activity.

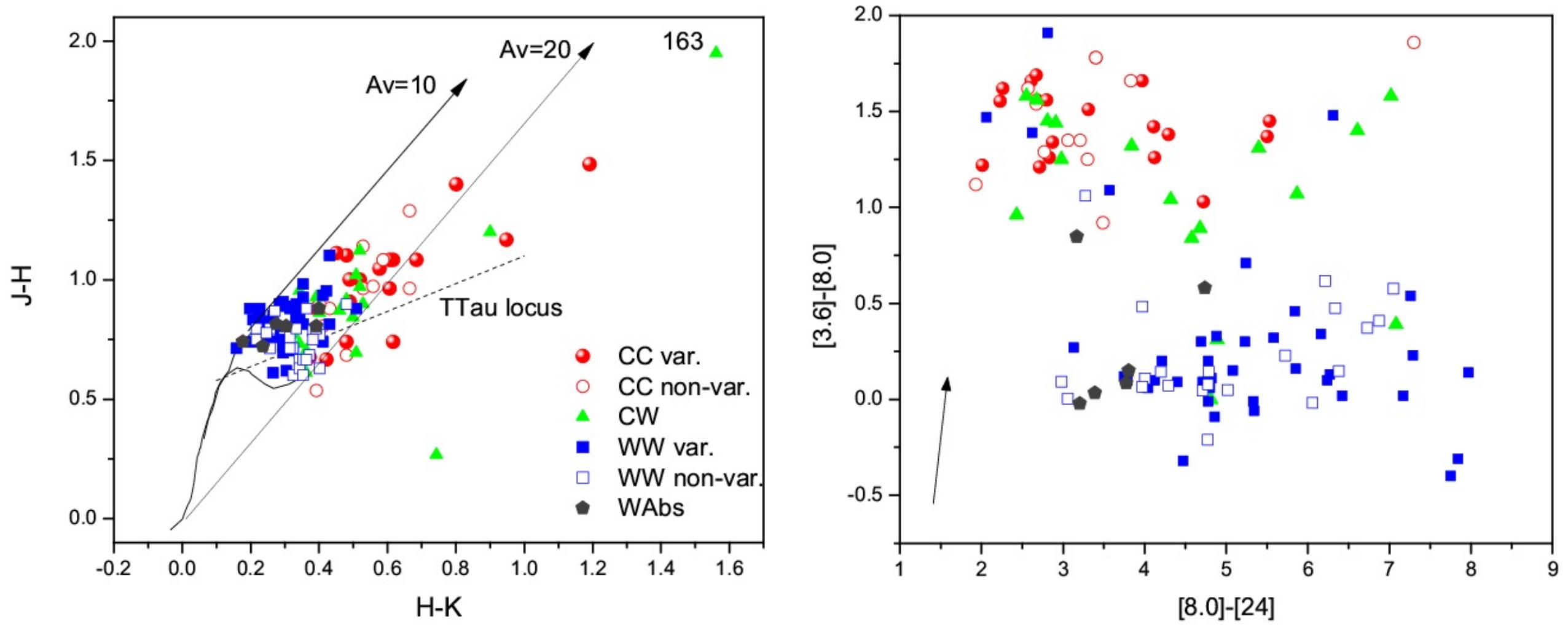

**Figure 4.** Two-colour diagrams of the stellar objects with $H_\alpha$ emission. In the left panel, the J-H vs. H-K diagram is presented. The dwarf and giant loci are taken from Bessell & Brett (1988) and converted to the CIT system (Carpenter, 2001). The arrows represent the reddening vectors (Rieke & Lebofsky, 1985). The locus of classical TTau stars is taken from Meyer et al. (1997). In the right panel, the [3.6]-[8.0] vs. [8.0]-[24] diagram is presented. The arrow represents the reddening vectors (Flaherty et al., 2007). A detailed explanation is in the text.

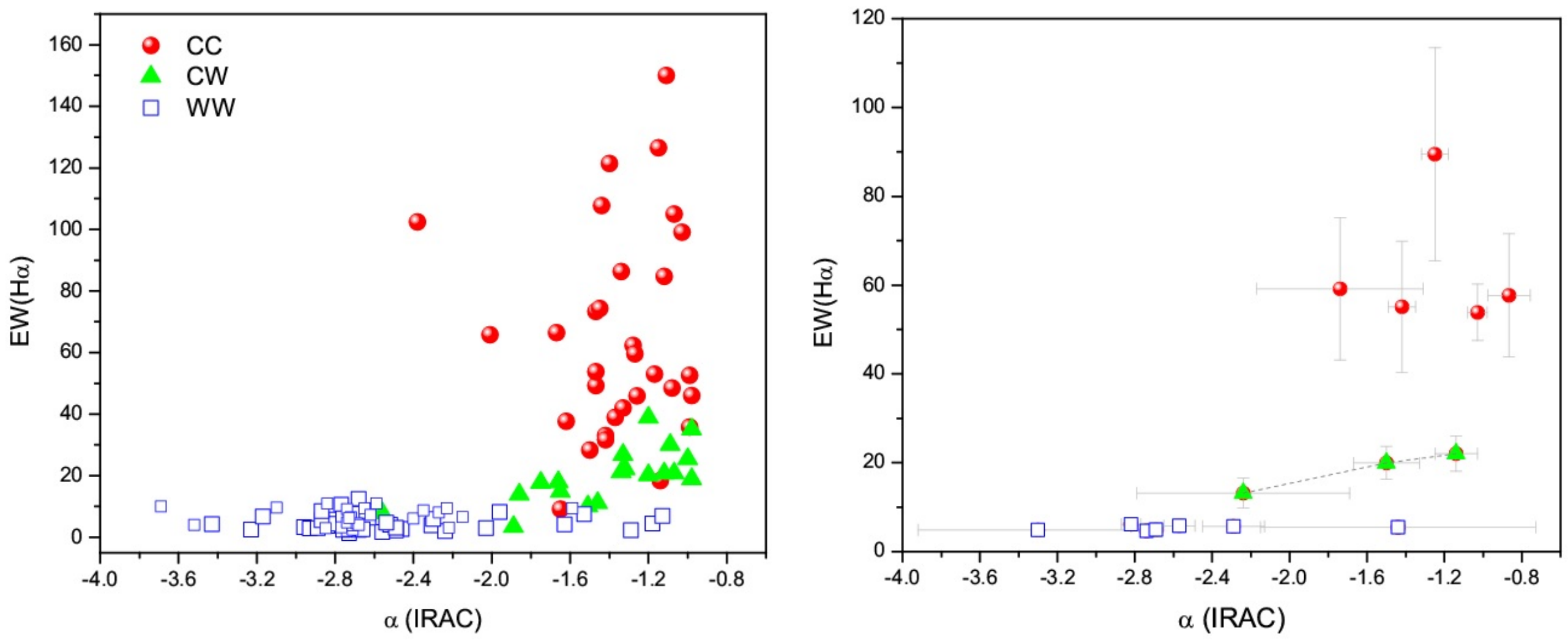

**Figure 5.** EW($H_\alpha$) vs. $\alpha^{3-8\,\mu\mathrm{m}}$ slope of SED. The binning in the right panel is the same as in Fig 3. The CC and WW samples include the both variable and non-variable objects. The dashed line is the result of the linear fitting of objects from the CW sample. There is a correlation between the EW ($H_\alpha$) and $\alpha^{3-8\,\mu\mathrm{m}}$ slope for CW sample but not for the CC and WW samples.

**Table 5** Masses, evolutionary ages and accretion rates

| N1 (1) | N2 (2) | Class. (3) | $V_E$ (4) | $M_*(M_{sun})$ (5) | Log (Age) (6) | $M_*$ ($M_{sun}$) (7) | Log (Age) (8) | Log ($\dot{M}$) (9) |
|---|---|---|---|---|---|---|---|---|
| 5 | 5 | CC | y | 2.2 | 6.6 | 2.5 | 6.6 | -7.3 |
| 29 | 29 | CC | y | 1.9 | 6.8 | 0.8 | 7.0 | -8.5 |
| 30 | 30 | CC | y | 0.7 | 6.1 | 1.4 | 6.5 | -8.1 |
| 37 | 37 | CC | y | 0.7 | 6.3 | 0.9 | 6.3 | -8.4 |
| 38 | 38 | CC | y | 0.8 | 6.4 | 1.1 | 6.4 | -7.7 |
| 39 | 39 | CC | y | 0.3 | 6.2 | 0.9 | 6.3 | -8.3 |
| 40 | 40 | CC | y | 0.4 | 6.4 | | | |
| 55 | 56 | CC | y | | | 0.9 | 6.7 | -7.5 |
| 59 | 59 | CC | y | 0.4 | 6.1 | 0.9 | 6.4 | -7.5 |
| 91 | 92 | CC | y | 0.5 | 6.5 | 0.5 | 6.6 | -8.2 |
| 93 | 94 | CC | y | 0.4 | 6.3 | 0.9 | 6.5 | -7.7 |
| 96 | 97 | CC | y | 0.4 | 6.3 | 0.4 | 6.2 | -8.6 |
| 108 | 110 | CC | y | 0.4 | 6.3 | 0.4 | 6.6 | -9.6 |
| 130 | 133 | CC | y | 0.3 | 6.7 | 0.8 | 6.5 | -7.9 |
| 138 | 142 | CC | y | 0.2 | 7.6 | 0.5 | 6.0 | -7.9 |
| 176 | 186 | CC | y | 0.5 | 7.3 | 0.9 | 6.1 | -8.1 |
| 207 | 228 | CC | y | 0.1 | 6.8 | 0.0 | 4.4 | -9.2 |
| 223 | 246 | CC | y | 0.1 | 6.8 | 0.2 | 5.8 | -8.0 |
| 247 | 276 | CC | y | | | 0.2 | 6.5 | -9.5 |
| 251 | 280 | CC | y | | | 1.1 | 5.5 | -7.4 |
| 15 | 15 | CC | n | 0.3 | 6.3 | 1.2 | 6.3 | -7.9 |
| 56 | 55 | CC | n | 0.6 | 6.6 | 0.3 | 5.6 | -7.5 |
| 57 | 57 | CC | n | 0.7 | 6.5 | 0.2 | 6.3 | -8.8 |
| 71 | 72 | CC | n | 0.5 | 6.3 | 0.8 | 6.5 | -8.2 |
| 116 | 119 | CC | n | 0.2 | 6.4 | 0.2 | 6.3 | -9.3 |
| 127 | 130 | CC | n | 0.2 | 6.4 | 0.3 | 6.4 | -9.2 |
| 142 | 146 | CC | n | 0.2 | 7.1 | 0.2 | 6.4 | -9.3 |
| 159 | 165 | CC | n | 0.2 | 6.6 | 0.3 | 6.6 | -9.4 |
| 172 | 182 | CC | n | 0.2 | 6.7 | 0.7 | 5.4 | -7.5 |
| 181 | 194 | CC | n | 0.2 | 6.6 | 0.3 | 6.8 | -10.1 |
| 190 | 204 | CC | n | 0.1 | 6.6 | 0.2 | 6.7 | -9.4 |
| | 113 | CC | n | 0.2 | 7.0 | | | |
| 20 | 20 | CW | y | 1.8 | 6.8 | 1.2 | 6.1 | -8.6 |
| 24 | 24 | CW | y | 0.6 | 6.2 | 0.2 | 3.9 | -6.9 |
| 34 | 34 | CW | y | 0.9 | 6.4 | 1.2 | 6.5 | -8.1 |
| 53 | 53 | CW | y | 0.5 | 6.1 | 0.5 | 6.2 | -8.4 |
| 81 | 82 | CW | y | 0.4 | 6.3 | 0.6 | 5.9 | -8.0 |
| 102 | 104 | CW | y | 0.4 | 6.3 | 0.4 | 6.6 | -9.0 |
| 111 | 114 | CW | y | 0.2 | 6.4 | 0.3 | 6.2 | -9.0 |
| 124 | 127 | CW | y | 0.2 | 6.4 | 0.4 | 5.4 | -9.0 |
| 129 | 132 | CW | y | 0.2 | 6.4 | 0.2 | 6.5 | -9.2 |
| 135 | 139 | CW | y | 0.2 | 6.4 | 0.1 | 6.7 | -9.6 |
| 147 | 151 | CW | y | 0.3 | 6.7 | 0.3 | 6.7 | -9.1 |
| 158 | 164 | CW | y | 0.2 | 6.6 | 0.2 | 6.1 | -7.8 |
| 163 | 171 | CW | y | | | 1.1 | 5.7 | -8.1 |
| 167 | 176 | CW | y | 0.2 | 6.7 | 0.2 | 6.5 | -10.6 |
| 179 | 192 | CW | y | 0.2 | 6.7 | 0.2 | 6.6 | -9.5 |
| 202 | 222 | CW | y | | | 0.2 | 6.7 | -9.5 |
| 238 | 264 | CW | y | 0.1 | 6.4 | 0.1 | 6.9 | -9.6 |
| 271 | 304 | CW | y | 0.2 | 7.0 | 0.4 | 5.6 | -8.1 |
| M185 | | CW | y | 0.3 | 6.6 | | | |
| | 103 | CW | y | | | | | |
| 31 | 31 | WW | y | 0.4 | 5.9 | 0.9 | 6.5 | -8.4 |
| 33 | 33 | WW | y | 0.9 | 6.2 | 0.9 | 6.4 | -8.3 |
| 47 | 47 | WW | y | 1.4 | 6.7 | 0.7 | 6.3 | -8.2 |
| 49 | 49 | WW | y | 0.5 | 6.0 | 0.8 | 6.4 | -8.5 |
| 61 | 61 | WW | y | 0.6 | 6.3 | 0.7 | 6.6 | -9.8 |
| 66 | 67 | WW | y | 0.2 | 6.4 | 0.4 | 6.2 | -8.4 |
| 68 | 69 | WW | y | 0.4 | 6.1 | 0.4 | 6.3 | -8.1 |
| 75 | 76 | WW | y | 0.5 | 6.2 | 0.6 | 5.6 | -6.9 |
| 76 | 77 | WW | y | 0.3 | 6.3 | 0.3 | 6.3 | -8.9 |
| 77 | 78 | WW | y | | | 0.4 | 6.4 | -8.8 |
| 78 | 79 | WW | y | 0.5 | 6.5 | 0.5 | 6.6 | -9.3 |
| 79 | 80 | WW | y | 0.3 | 6.2 | 0.5 | 6.4 | -7.8 |
| 80 | 81 | WW | y | 0.3 | 6.9 | 0.4 | 5.9 | -7.8 |
| 82 | 83 | WW | y | 0.4 | 6.2 | 0.4 | 6.4 | -8.9 |
| 83 | 84 | WW | y | 0.4 | 6.3 | 0.4 | 6.5 | -9.1 |
| 85 | 86 | WW | y | 0.2 | 6.5 | 0.3 | 6.3 | -8.8 |

**Table 5 – continued from previous page**

| N1 (1) | N2 (2) | Class. (3) | $V_E$ (4) | $M_*(M_{sun})$ (5) | Log (Age) (6) | $M_*$ $(M_{sun})$ (7) | Log (Age) (8) | Log $(\dot{M})$ (9) |
|---|---|---|---|---|---|---|---|---|
| 87 | 88 | WW | y | 0.4 | 6.1 | 0.3 | 5.9 | -9.6 |
| 88 | 89 | WW | y | 0.2 | 6.6 | 0.3 | 6.3 | -9.1 |
| 89 | 90 | WW | y | 0.3 | 6.3 | 0.3 | 6.4 | -8.9 |
| 94 | 95 | WW | y | 0.6 | 6.5 | 0.5 | 6.7 | -9.0 |
| 95 | 96 | WW | y | 0.3 | 6.4 | 0.2 | 4.6 | -7.4 |
| 99 | 100 | WW | y | 0.9 | 7.2 | 0.6 | 5.4 | -9.2 |
| 101 | 102 | WW | y | 0.4 | 6.5 | 0.4 | 6.6 | -9.0 |
| 103 | 105 | WW | y | 0.4 | 6.5 | 0.4 | 6.6 | -8.9 |
| 105 | | WW | y | 0.2 | 7.0 | 0.4 | 6.6 | -9.0 |
| 107 | 109 | WW | y | 0.3 | 6.3 | 0.4 | 6.6 | -9.1 |
| 109 | 111 | WW | y | 0.4 | 6.4 | 0.4 | 6.5 | -9.0 |
| 110 | 112 | WW | y | 0.2 | 6.4 | 0.3 | 6.2 | -8.3 |
| 118 | 121 | WW | y | 0.5 | 6.5 | 0.4 | 6.5 | -7.8 |
| 120 | 123 | WW | y | 0.3 | 7.0 | 0.5 | 6.8 | -9.0 |
| 121 | 124 | WW | y | 0.2 | 6.4 | 0.2 | 5.2 | -7.5 |
| 122 | 125 | WW | y | 0.5 | 6.8 | 0.5 | 6.8 | -8.2 |
| 128 | 131 | WW | y | 0.2 | 6.4 | 0.4 | 6.5 | -8.0 |
| 134 | 138 | WW | y | 0.2 | 6.4 | 0.3 | 6.5 | -9.5 |
| 141 | 145 | WW | y | 0.3 | 6.5 | 0.3 | 6.7 | -9.4 |
| 148 | 153 | WW | y | 0.3 | 6.6 | 0.3 | 6.7 | -9.7 |
| 160 | 166 | WW | y | | | 0.3 | 6.6 | -9.3 |
| 166 | 174 | WW | y | 0.3 | 6.6 | 0.3 | 6.8 | -9.7 |
| 169 | 179 | WW | y | 0.2 | 6.6 | 0.3 | 6.7 | -9.8 |
| 173 | 183 | WW | y | 0.2 | 6.7 | 0.3 | 6.8 | -9.0 |
| 186 | 200 | WW | y | 0.2 | 6.7 | 0.3 | 5.4 | -9.2 |
| 195 | 211 | WW | y | 0.2 | 6.7 | 1.8 | 6.7 | -12.8 |
| 209 | 230 | WW | y | 0.2 | 6.7 | 0.3 | 4.9 | -8.1 |
| 233 | 256 | WW | y | 0.1 | 7.1 | 0.3 | 5.4 | -7.7 |
| 45 | 45 | WW | n | 1.0 | 6.5 | 0.7 | 6.5 | -8.6 |
| 58 | 58 | WW | n | 0.2 | 6.6 | 0.3 | 6.1 | -10.3 |
| 60 | 60 | WW | n | 0.5 | 6.4 | 0.5 | 6.4 | -9.8 |
| 65 | 65 | WW | n | 0.5 | 6.3 | 0.5 | 6.5 | -7.9 |
| 98 | 99 | WW | n | 0.2 | 6.3 | 0.3 | 6.3 | -9.0 |
| 119 | 122 | WW | n | 0.6 | 6.7 | 0.4 | 6.7 | -8.9 |
| 139 | 143 | WW | n | 0.2 | 6.4 | 0.3 | 6.3 | -7.9 |
| 143 | 147 | WW | n | 0.4 | 6.7 | 0.4 | 6.8 | -9.4 |
| 150 | 156 | WW | n | 0.3 | 6.4 | 0.3 | 6.6 | -9.4 |
| 152 | 158 | WW | n | 0.2 | 6.5 | 0.3 | 6.7 | -9.2 |
| 154 | 160 | WW | n | 0.3 | 6.7 | 0.3 | 6.7 | -9.4 |
| 162 | 169 | WW | n | 0.2 | 6.6 | 0.3 | 6.7 | -9.6 |
| 168 | 177 | WW | n | 0.2 | 6.6 | 0.2 | 6.7 | -10.1 |
| 170 | 180 | WW | n | 0.2 | 6.6 | 0.2 | 6.2 | -8.3 |
| 171 | 181 | WW | n | 0.2 | 6.6 | 0.3 | 6.6 | -9.6 |
| 175 | 185 | WW | n | 0.1 | 6.6 | 0.3 | 5.2 | -9.2 |
| 187 | 201 | WW | n | | | 0.3 | 6.8 | -9.8 |
| 189 | 203 | WW | n | 0.1 | 6.8 | 0.2 | 5.0 | -8.4 |
| 197 | 216 | WW | n | 0.1 | 6.7 | 0.3 | 4.9 | -7.5 |
| 198 | 217 | WW | n | 0.1 | 6.7 | 0.2 | 6.8 | -8.9 |
| 206 | 227 | WW | n | | | 0.2 | 5.0 | -8.3 |
| 212 | 233 | WW | n | 0.3 | 6.6 | 0.2 | 4.9 | -7.4 |
| 227 | 250 | WW | n | 0.1 | 6.9 | 0.3 | 5.4 | -9.3 |
| 228 | 251 | WW | n | | | 0.2 | 6.6 | -8.5 |
| 232 | 255 | WW | n | 0.1 | 6.2 | 0.1 | 6.9 | -9.4 |
| 23 | 23 | WAbs | | 0.8 | 6.1 | 1.3 | 6.4 | -9.2 |
| 32 | 32 | WAbs | | 1.8 | 6.3 | 1.0 | 6.1 | -10.5 |
| 52 | 52 | WAbs | | 0.4 | 6.5 | 1.0 | 7.0 | -12.5 |
| 62 | 62 | WAbs | | 0.8 | 6.5 | 1.0 | 6.7 | -11.1 |
| 126 | 129 | WAbs | | 0.4 | 6.4 | 0.5 | 6.7 | -9.7 |
| 149 | 155 | WAbs | | 0.2 | 6.5 | 0.3 | 6.5 | -10.4 |

This reflects the data in Table 3, which gives for each sample the number of stellar objects considered in Flaherty et al. (2013) and the number of variables among them. The current model for this process is magnetospheric accretion where the stellar magnetic field threads the disc, and the material in the disc falls along the field lines onto the surface of the star (Koenigl, 1991). The gas flows along the stellar magnetic fields lines and creates a shock as it strikes the star surface, which leads to the formation of hot spots. According to the explanation in Flaherty et al. (2013), the variable illumination from the hot spots is one of the possible mechanisms for explaining the observed infrared variability. On the other hand, the emission from this shocked-material is strongest at shorter wavelengths, including the line emission in the optical range. This would explain the correspondence between MIR variability and $H_\alpha$ activity.

Concerning the colour index [8.0] - [24] (see Fig. 4 and Table 3), the opposite situation is observed: with increasing $H_\alpha$ activity the infrared excess decreases. The objects from the WW sample have the largest colour index. A little or no near-infrared excess but large mid-infrared excess are distinctive for a subset of PMS objects called "transitional disk" systems. The nature of transition disks is controversial, but in general, it is assumed that these stellar objects may be dust free, therefore optically thin, within a certain radius close to the central star. On the other hand, a significant mid-infrared excess observed in these stellar objects indicates the presence of a relatively massive outer disk (Hartmann, 2009, and ref. therein). That is why, the 24 $\mu$m band of the MIPS camera of the *Spitzer* telescope is very productive for diagnosing stars of the III evolutionary class (Whitney et al., 2003; Muzerolle et al., 2004; Wolk et al., 2010).

### *3.2.3 X-ray data*

X-ray data plays a critical role in studies of different aspects of the star-forming process. These include solar-like coronal/magnetospheric flares, an accretion funnel flow, an outflow, and other manifestations of stellar activity on different evolutionary stages. Therefore, X-ray radiation is detected in both Class II (CTT) and Class III (WTT) objects. Most studies found that CTT objects have systematically lower X-ray luminosities than WTT (Feigelson et al., 2007; Stelzer et al., 2012; Flaccomio et al., 2003). On the other hand, it is known that X-rays are very sensitive to extinction (Wolk et al., 2006). X-ray radiation in CTT objects is mainly created as a result of accretion activity that in turn assumes the presence of a massive circumstellar disk, which can absorb short-wave radiation substantially. Presumably, this explains the relatively fainter X-ray luminosity in CTT objects.

To test this assumption, we have considered the dependence of X-ray luminosity on extinction ($A_v$). We took the X-ray luminosity from Preibisch & Zinnecker (2002) and Stelzer et al. (2012) and the values of $A_v$ from Luhman et al. (2003). It should be noted that X-ray data are determined for 93% of the total sample of the objects. In Fig. 6 the diagrams of $\log(L_x/L_{bol})$ versus extinction ($A_v$) are presented. As in the previous cases, for greater visibility we have split the stellar objects from different sample into several bins with a nearly equal numbers. The diagrams clearly show that there is a well-defined inverse relationship between these two parameters. The fraction of X-ray radiation increases from the early evolutionary classes to the later ones. This also reflects the averaged values of $\log(L_x/L_{bol})$ for different samples, presented in Table 3. Among the stellar objects from the CC sample there is a barely difference between variable and non-variable objects. The average X-ray flux of the non-variables is slightly greater that, to some extent, can be explained by a smaller value of $<A_v>$. In contrast, in the WW sample with a lower value of $A_v$, the X-ray activity of the non-variable objects is slightly smaller. In this case, the explanation could be lower flare activity. The objects from the CW sample occupy an intermediate position with respect to both $\log(L_x/L_{bol})$ and $A_v$. The objects from sample WAbs have the lowest fraction of X-ray radiation. Apparently, this is due to the fact that these objects are at a later evolutionary stage and their flare activity is low.

Of course, it should be noted that large values of standard deviations of the median values mean that the difference between the samples is really "barely significant". But these differences well consistent with the expectations of the evolutionary stages of the various samples.

### *3.2.4 Masses, evolutionary ages and accretion rates*

Two methods were used to determine the masses and the evolutionary ages of the $H_\alpha$ emitters. First, we used the isochrone models adopted from Siess et al. (2000). The position of the objects relative to the isochrones is shown on the HR diagram in Fig. 7. The values of the masses and the ages of the stellar objects from our sample are presented in Table 5 (columns (5) and (6)). Both medians of the ages and the averages of masses for every sample are presented in Table 3. The arrangement of the stellar objects relative the isochrones, as well as the median values in Table 3 confirm the results obtained in previous studies that the mean age of the cluster is $2-3$ Myr (Herbst, 2008, and ref. herein). Furthermore, there is no noticeable difference between the objects with different $H_\alpha$ activity, as has already been pointed out in Herbig (1998). Nevertheless, we want to note that the median values of the variable objects in the CC and WW samples are slightly less those that of the non-variable ones. The difference between the variable and non-variable objects is much more pronounced relative to their masses (see Table 3). In both samples (CC and WW) the averaged values of the masses of the variables are noticeably higher.

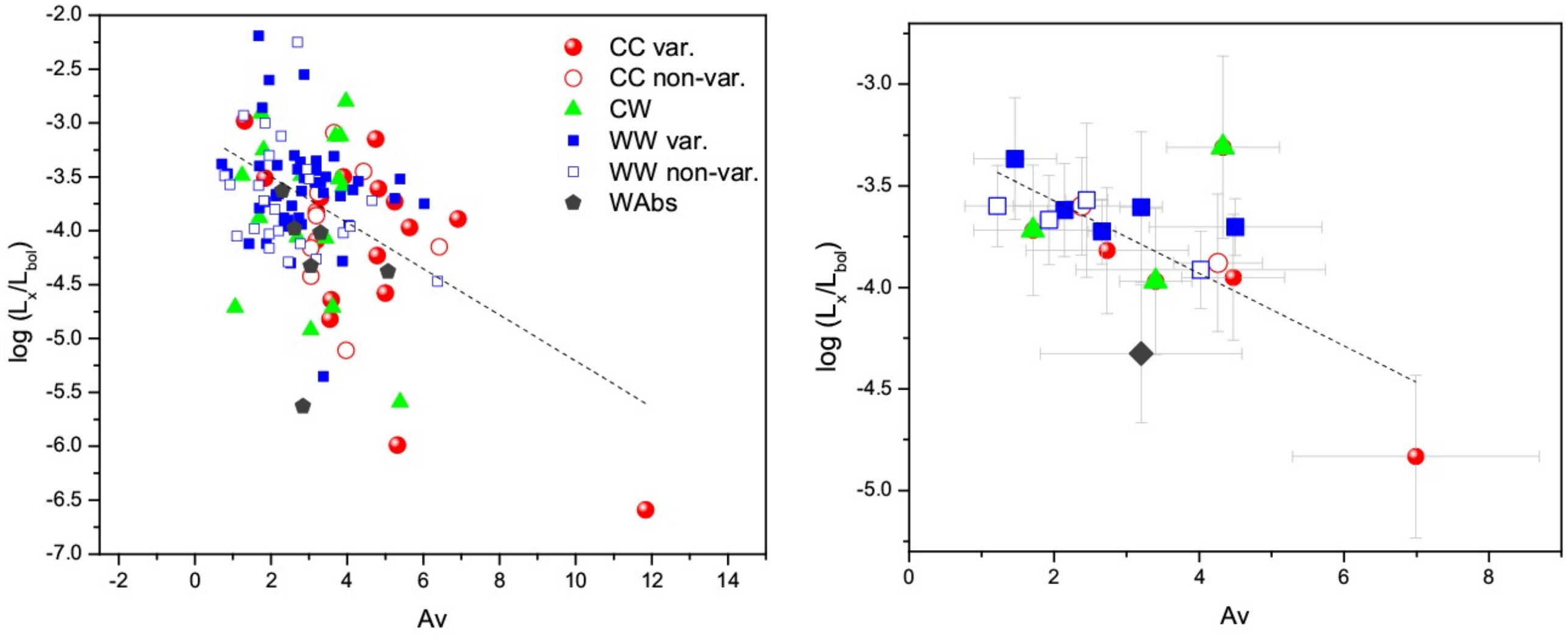

**Figure 6.** Log($L_X/L_{bol}$) vs. extinction ($A_v$). The binning on the right panel are the same as in Fig 3. The linear fitting of all objects (the dashed lines in both panels) clearly shows that there is a well-defined inverse relationship between these two parameters.

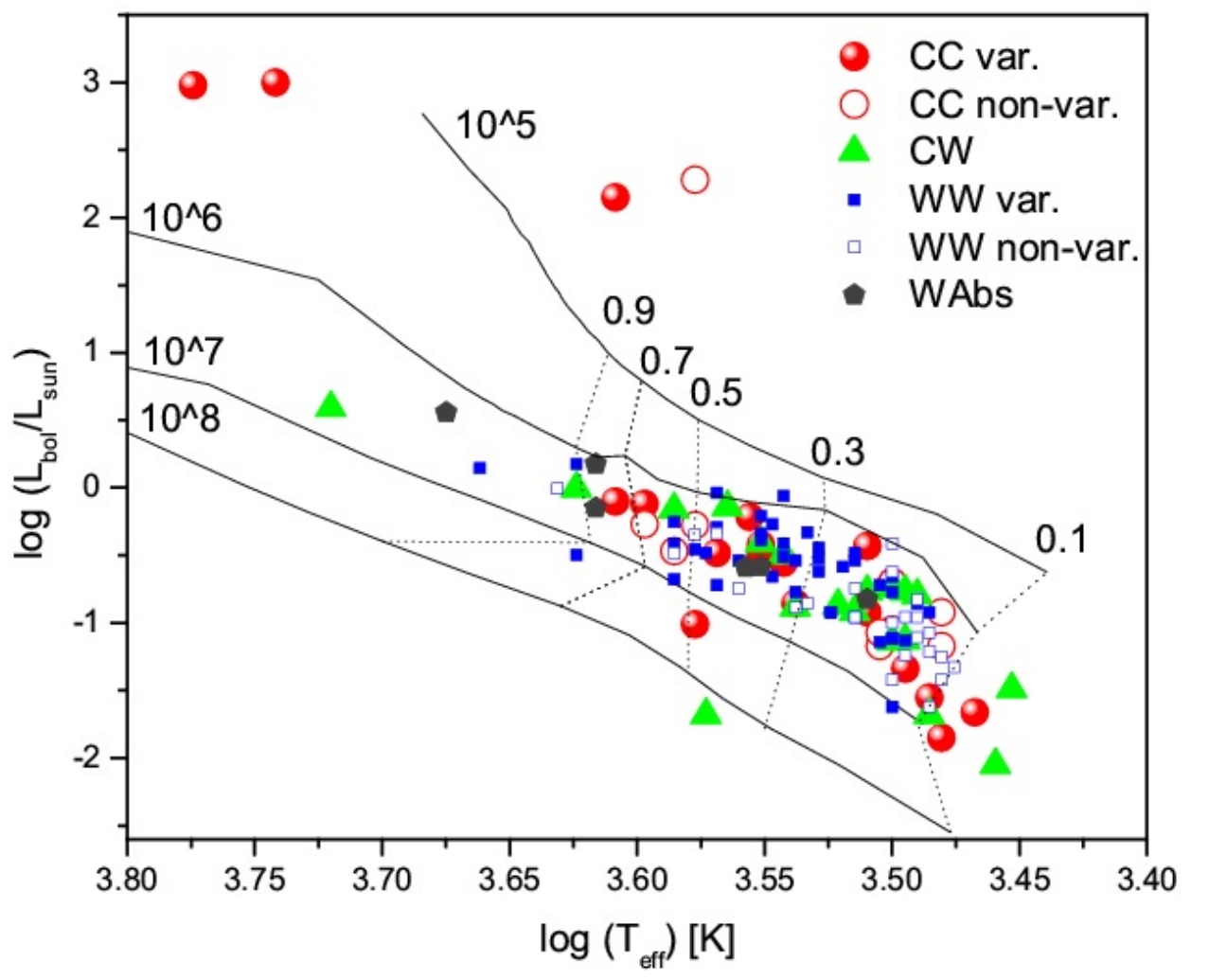

**Figure 7.** HR diagram for $H_\alpha$ emitters. Mass tracks and isochrones are adopted from Siess et al. (2000). There is no noticeable difference in the location of objects with different $H_\alpha$ activity relative to the isochrones.

The stars in the CW sample occupy an intermediate position. The most massive objects belong to the sample WAbs.

We also determined the masses, the evolutionary ages, as well as the mass accretion rate for the $H_\alpha$ emitters using the SED fitting tool of Robitaille et al. (2007). We have used the command-line version of the SED fitting tool, where a large number of precomputed models are available. For the SED fitting, we used the averaged over all epochs of observations R and I magnitudes, 2MASS J, H, K magnitudes, as well as the photometric data of the [3.6], [4.5], [5.8], [8.0] and [24] bands (see data in Table 4). The SED fitting tool might not be able to give distinct information for variable sources. To minimize this problem, a 10% uncertainty was assumed for each band. We fixed the distance range from 280 to 320 pc (Herbst, 2008). The $A_v$ range was fixed from minimal to maximal values for every sample. To identify the representative values of different physical parameters, the tool retrieved all models, for which the differences between their $\chi^2$ values and the best $\chi^2$ were smaller than 3N, where N is the number of the data points (as suggested in Robitaille et al., 2007). This approach was taken because the sampling of the model grids is too sparse to effectively determine the minima of the $\chi^2$ surface and consequently obtain the confidence intervals. In Table 5 the weighted averages and the standard deviations of the masses, the ages, and the mass accretion rates corresponding to the models with $\chi^2$ - $\chi^2_{\rm best}$ < 3N for all objects from our sample are given (columns (7) - (9)). The average values of these parameters for each sample are presented in Table 3. As a result of the application of the SED fitting tool, the estimation of the masses of objects with the early evolutionary stage (sample CC), both variables and non-variables, increased. The average evolutionary age of objects from all samples, excluding WAbs, does not show a tangible difference. The evolutionary age of the objects from WAas samples has increased.

The mass accretion rates ranged from a few $10^{-11}$ to $10^{-7}$ $M_{\rm sun}$/yr. The average values of the log $\dot{M}$ and standard deviations for the samples are presented in Table 3. The variables from the CC sample have the highest value, and the objects from the WAbs sample have the smallest one. For the remaining samples, including non-variables from the CC sample, the mass accretion rate is practically the same. It should be noted that in the WW sample the average value of log$\dot{M}$ of the variables is slightly higher than that of the non-variables.

For several objects from our sample (N1 = 5, 31, 37, 32, 61 in Tables 3 and 4), the values of log $\dot{M}$ were determined on the basis of the spectral data (Dahm, 2008). We compared the results with our data. In the first three cases (5, 31 and 37) the obtained values coincide with an accuracy of 0.1. For the other two objects (32 and 61), however, the log $\dot{M}$ values given in Dahm (2008) exceed our data ∽1.5 times. Let us note that the second object (N1 = 61) belongs to the sample with the lowest $H_\alpha$ activity (WAbs).

In Fig. 8 the mass accretion rates as a function of the stellar mass (determined using the SED fitting tool) are presented. It is clearly seen that for objects from the samples CC and CW, there is a clear correlation between log $\dot{M}$ and the stellar mass: with mass increasing the mass accretion rate also increases. The slope and R-square of linear fitting are equal to 1.51 ± 0.28 and 0.37, respectively. The statistical slope 1.51 ± 0.28, which corresponds to a power-law $M_{acc} \propto M^{1.51 \pm 0.28}$, is within the error bar in agreement with the result obtained for the young stellar cluster NGC 2264 for the same range of the stellar masses (Venuti et al., 2014). The evolutionary ages of the clusters NGC 2264 and IC 348 are almost identical. It should be noted that the mass accretion in NGC 2264 was determined based on U-band photometry. For objects from the sample WW, there is no relationship between the mass accretion rate and stellar masses. Most of the stars from the CW sample (75%) on the basis of the $\alpha$ slope of the SEDs can be classified as objects of the earlier II evolutionary class (CTT objects), in which $H_\alpha$ emission is mainly formed as a result of accretion activity, rather than chromospheric flares, like in objects of the III evolutionary class (WTT objects). Apparently, therefore, the stars of this sample also show a relationship between these two parameters.

We also considered the mass accretion rates as a function of the evolutionary age (see Fig. 9). We used the data obtained using the SED fitting tool (Robitaille et al., 2007). As in the previous cases, for greater visibility we split the stellar objects from different sample into several bins with a nearly equal numbers. According to the diagrams in Fig. 9, there is a certain dependence between the mass accretion rates and the stellar ages (the slope and R-square of linear fitting are equal to -1.23 ± 0.21 and 0.77, respectively). The statistical slope -1.23 ± 0.21, which corresponds to power-law $M_{acc} \propto M^{-1.23 \pm 0.21}$, is somewhat more than the value of the slope obtained for the young stellar cluster NGC 2264 (Venuti et al., 2014). In general, the accretion activity of objects decreases with age. Nevertheless, objects from different samples manifest themselves in different ways. According to the averaged data (Fig. 9, right panel), the log$\dot{M}$ of the variables from the CC sample, which have a significant age dispersion, remains practically at the same level (the slope and R-square of linear fitting are equal to -0.03 ± 0.35 and 0.98, respectively). In Manara et al. (2012) it was found that in Orion nebula cluster there are different accretion evolutionary timescales for sources with different mass. The mass accretion rates of more massive stellar objects decay more slowly. Recall that according to both the evolutionary model of Siess et al. (2000) and the SED fitting tool (Robitaille et al., 2007), the variables from the CC sample are more massive than the objects from the other samples, excluding the WAbs. Consequently, the finding made in Manara et al. (2012) is also applicable for objects in the IC 348 cluster.

It should also be noted that there is a group of objects from the WW sample with the lowest evolutionary age and the highest mass accretion rate. It cannot be excluded that this is a result of measurement errors, which can lead to the misclassification based on EW($H_\alpha$). It could also be the result of certain approximations in the modelling. In particular, for the SED fitting tool, we used the averaged data of photometry obtained at different epochs of observation, although the overwhelming majority of objects are variable. All this, of course, can lead to certain distortions of the real representation.

## 4 DISCUSSION

The results of the optical observations of the young stellar cluster IC 348 confirmed the observations made earlier that in this star-forming region, there is a significant number of stellar objects with variable $H_\alpha$ emission (Herbig, 1998; Stelzer et al., 2012). In total, from 127 examined stellar objects, 90 show significant variability of EW($H_\alpha$). In addition, we found a group of stars (20 objects), which showed not only a significant variability of the equivalent width, but also an apparent change in their evolutionary stage (CTT⇆WTT).

The analysis of the data, which cover a wide spectral range (from X-ray to mid-infrared) has shown that there are certain differences not only between the stellar objects with different stages of $H_\alpha$ activity (CTT and WTT), but also between the stars with variable and non-variable EW($H_\alpha$). With only slight difference in their evolutionary ages, the variables in both samples in a number of parameters, including mass accretion rate and X-ray radiation, are more active.

According to the current model, the photometric and spectral variability of PMS low-mass stars is mainly due to the magnetospheric accretion (CTT) or chromospheric magnetic flares (WTT). Indeed, several studies, including extensive photometric monitoring and spectropolarimetric observations of individual objects (e.g., Donati et al., 2010), have shown that the surface of T Tau objects are possibly covered by multiple hot (accretion) or cold (chromospheric flares) spots. The area covered by spots can be deduced from the observed vari-ability amplitudes at different wavelengths, including $H_\alpha$ emission. As stated in Venuti et al. (2015), for both accreting and non-accreting stars, the medium-term rotational modulation with hot and cold spots is the leading time scale for variability up to several years, which was shown by our observations, at least for 70% of T Tau objects. This fully explains the agreement between the measure of variability of $H_\alpha$ emission and the activity of stellar objects with respect to other parameters.

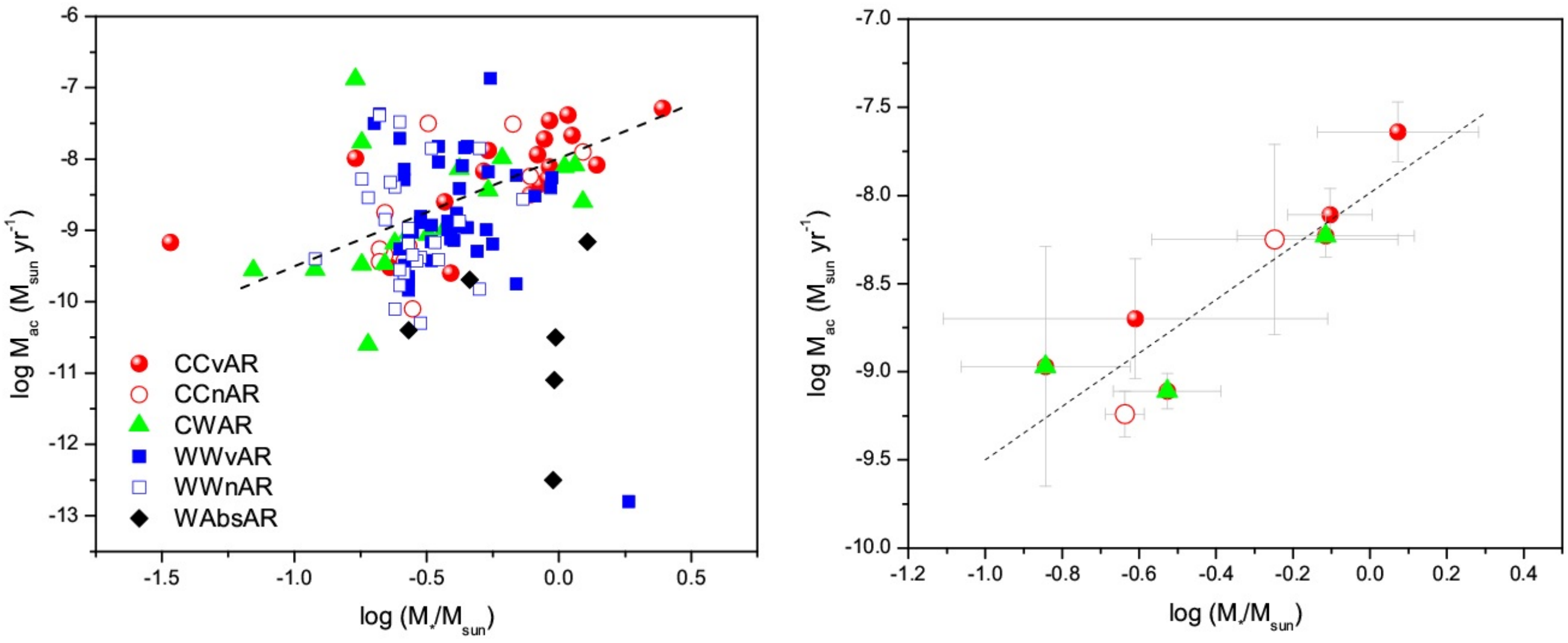

**Figure 8.** Mass accretion rates vs. stellar masses (determined using the SED fitting tool). The binning on the right panel are the same as in Fig 3. The dotted line is the result of the linear fitting for CC and CW samples. The fitting shows that there is a correlation between the mass accretion rate and stellar masses for CC and CW objects but not for the WW and Wabs samples.

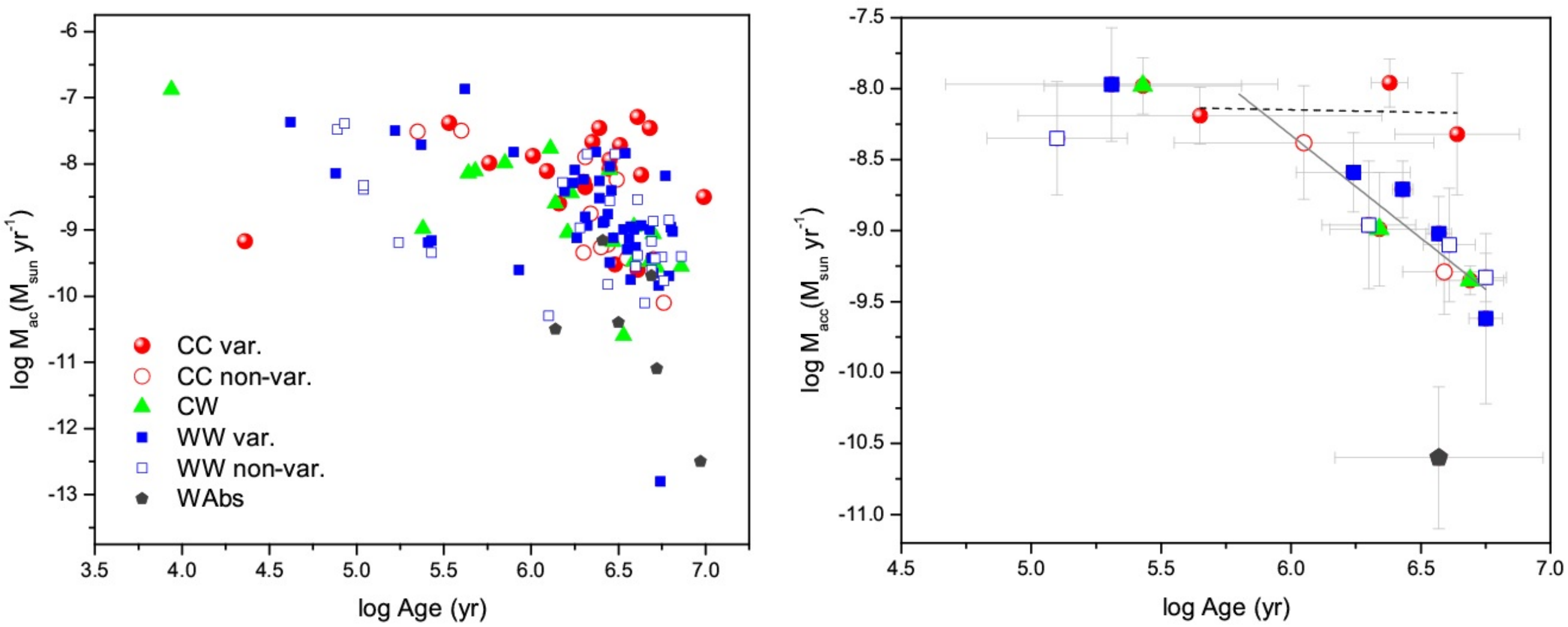

**Figure 9.** Mass accretion rates vs. evolutionary age (determined using the SED fitting tool). The binning in the right panel are the same as in Fig 3. The dashed line is the result of linear fitting of CC variable objects and the solid line is the result of the linear fitting of other objects. There is a certain dependence between the mass accretion rates and the SED fitting evolutionary age for all objects except the CC variables. The log $\dot{M}$ of the variables from the CC sample practically does not depend on the evolutionary age.

The other distinct difference between the variable and non-variable objects relative to the EW($H_\alpha$) is the average values of their masses. The variables from both CTT and WTT samples are noticeably more massive than non-variables.

As was shown in a number of works devoted to the study of the accretion process (e.g., Muzerolle et al., 2003; Hartmann et al., 2006; Rigliaco et al., 2011), two relationships $M_{\rm acc} \propto {M_*}^b$ and $M_{acc} \propto t^{-\eta}$ likely reflect the actual mechanisms governing the accretion process. Indeed, more massive variable CC objects have a higher accreting rate, which agree with $M_{\rm acc} \propto {M_*}^b$ relationship. On the other hand, as shown on Fig. 9, the accretion rate of these stellar objects is practically independent of evolutionary age. Hartmann et al. (2006) reviewed a number of different mechanisms, which may be relevant to disk accretion and its variability, as well as to the $M_{\rm acc} \propto {M_*}^b$ relationship. As noted by those authors, several unrelated mechanisms may contribute to establishing this relationship, and those mechanisms may act differently in different mass regimes. Therefore, the values of the two power-law exponents ($\eta$ and b) may vary depending on the mass range (Manara et al., 2012). In a number of the previous studies it was shown that the decay of mass accretion rates during the evolutionary age of more massive stellar objects is slower (Venuti et al., 2014; Manara et al., 2012). Indeed, in our case, with a wide spread of age (log (Myr) = 5.35 - 6.76), the mass accretion rate of comparatively more massive CC variables, unlike other objects, remains almost unchanged. That confirms the conclusions made in previous works.

Apparently, a similar trend is also present in objects that are at a later stage of evolution (WW sample), whose $H_\alpha$ emission, according to the current model, are mainly due to chromospheric flares. WW variables are also more active and massive. This can be due the fact that cool spots that lead to variability of $H_\alpha$ emission, are more widespread and the probability of their detection is greater.

On the other hand, cases of WTT young stars that episodically cross the CTT-defining borders have been reported in the literature (Murphy et al., 2011; Cieza et al., 2013). Taking into account the fact that the averaged value of the mass accretion rate of WW variables does not practically differ from non-variable CC and CW objects, this scenario is quite possible. And, as in the case of the CC objects, the decline in the accretion activity occurs more slowly in more massive objects. This suggests that the variability of $H_\alpha$ emissions, at least in part, can be explained by residual accretion activity.

Nevertheless, in connection with these, it should be noted that, according to our estimations, the most massive are the most "passive" objects from the WAbs sample, in which the $H_\alpha$ line alternately is observed in emission and absorption. According to the model adopted from Robitaille et al. (2007), these objects are "older" than the rest, and they could be formed during an earlier star formation wave. According to Siess et al. (2000) model, however, their age is comparable with WTT objects. In all cases, the small number of these objects does not permit making definite assumptions.

Objects, which changed their apparent evolutionary state at different epochs of observations (CW sample), according to some parameters, including the infrared excess, the extinction, the X-ray intensity, and the mass accretion rate occupy an intermediate position between CC and WW variables. And they can be considered as the stars that are in a transitional evolutionary stage between CTT and WTT objects. Nevertheless, they have some differences: the averaged value of the variability fraction $f_v$ is significantly (~1.5 times) larger than in the CC and WW samples; the largest percentage (~80%) of the objects with periodical variability of optical brightness; the lack of dependence between the $H_\alpha$ variability fraction ($f_v$) and standard deviation of R magnitudes; the well-defined relationship between EW($H_\alpha$) and the $\alpha^{3-8\,\mu\rm m}$ slope of SEDs. We can assume that the variability of these stellar objects, at least in a certain fraction of them, is due to the fact that they are close, unresolved binaries with different stages of $H_\alpha$ activity (CTT and WTT), which affects and modulates their $H_\alpha$ emission.

## 5 SUMMARY

The photometric and slit-less observations at several epochs in the central region of the young stellar cluster IC 348 revealed a significant number of stars with variable $H_\alpha$ emission. In total, from 127 examined stellar objects, 90 show the significant variability of EW($H_\alpha$). We have defined evolutionary classes relative to $H_\alpha$ activity for all objects from our sample for all epochs of the observations. As a result, 32 were classified as CTT and 69 as WTT objects. The fraction of the objects with variable EW($H_\alpha$) among these two samples is almost the same and is ~60%. We also identified 20 stellar objects, which show not only a significant variability of the equivalent width, but also change the evolutionary stage (CTT⇆WTT). For 6 stars the $H_\alpha$ line was observed both in emission and in absorption.

The comparative analysis of variability of $H_\alpha$ emission with other parameters, including optical, infrared, and X-ray data has shown the following:

- The averaged value of the extinction (Av) decreases from CTT to WTT objects.
- There is a clearly expressed relationship between EW($H_\alpha$) and infrared excess: the J-K and [3.6]-[8.0] colour indices decrease significantly with decreasing $H_\alpha$ activity, while the [8.0]-[24] colour index has an inverse dependence on the EW($H_\alpha$).
- Except for the CW objects, the correspondence between the classification of the evolutionary stage according to the $\alpha^{3-8\,\mu m}$ slope and EW($H_\alpha$) reaches 90% and more.
- The intensity of X-ray radiation increases with decreasing EW($H_\alpha$). But, on the other hand, there is well-defined inverse relationship between Av and X-ray intensity.
- The evolutionary age of the CTT and WTT objects is almost the same.
- There is a correlation ($M_{acc} \propto M^{1.51\pm0.28}$) between the mass accretion rate and stellar masses for CTT objects and stars with variable evolutionary stage.
- There is a correlation ($M_{acc} \propto M^{-1.23\pm0.21}$) between the mass accretion rate and evolutionary age for non-variable CTT, CW, and WTT objects.

Data analysis also revealed the distinct differences between the variables and non-variables objects relative to the EW($H_\alpha$):

- The averaged values of Av are higher for the CTT and WTT variables than of non-variables.
- The masses of the CTT and WTT variables are noticeable higher than of non-variables.
- With a wide spread of age (log (Myr) = 5.35 - 6.76), the mass accretion rate of CTT variables, unlike other objects, remains almost unchanged.
- The WTT variables, like CTT ones, are more massive and more active relative to mass accretion rate and X-ray radiation, than non-variables.

Summarizing the obtained data, we infer that variables from both CTT and WTT samples are at an earlier evolutionary stage, than non-variable ones.

Objects, which changed their apparent evolutionary state at different epochs of observations, have some differences, including a larger value of the variability fraction $f_v$ (∼1.5 times) and the largest percentage (∼80%) of the objects with periodical variability of optical brightness. We can assume that the variability of these stellar objects, at least in a certain fraction of them, is due to the fact that they are close, unresolved binaries with different stages of $H_\alpha$ activity (CTT and WTT), which affects and modulates their $H_\alpha$ emission.

As noted above, the mean age of the members of IC 348 cluster ($2-3$ Myr) corresponds to the time where the structure of the disks of most young stellar objects changes from primordial, rather massive accretion disks to transitional and debris disks, and the point in time when planetary formation is thought to occur. Undoubtedly, it is interesting to conduct such studies and compare the data obtained in younger and older clusters and associations such e.g. Serpens cluster with age ∼1 Myr (Winston et al., 2005) or the Orion Nebula Cluster with age $1-2$ Myr (Hillenbrand, 1997) and e.g. Upper Scorpius with age ∼5 Myr (de Geus et al., 1989).

## 6 ACKNOWLEDGEMENTS

We are very grateful to the anonymous referee for helpful comments and suggestions. We thank the management of NAS RA Byurakan Astrophysical Observatory for providing the program of observations on 2.6 m telescope. This publication makes use of data products from the Two Micron All Sky Survey, which is a joint project of the University of Massachusetts and the Infrared Processing and Analysis Center/California Institute of Technology, funded by the National Aeronautics and Space Administration and the National Science Foundation. This work was made possible by a research grant from the Armenian National Science and Education Fund (ANSEF) based in New York, USA.